\documentclass[lettersize,journal]{IEEEtran}
\usepackage{amsmath,amsfonts}
\usepackage{algorithmic}
\usepackage{algorithm}
\usepackage{array}
\usepackage[caption=false,font=normalsize,labelfont=sf,textfont=sf]{subfig}
\usepackage{textcomp}
\usepackage{stfloats}
\usepackage{url}
\usepackage{verbatim}
\usepackage{graphicx}
\usepackage{cite}
\usepackage{xcolor}
\usepackage{hyperref}
\usepackage{booktabs}
\usepackage{multirow}
\usepackage{xcolor}
\usepackage{textcomp}
\usepackage{orcidlink}
\hypersetup{
	colorlinks=true,
	linkcolor=blue,     
	citecolor=green     
}

\begin{document}

\title{Mamba-FCS: Joint Spatio-\textbf{F}requency Feature Fusion, \textbf{C}hange-Guided Attention, and \textbf{S}eK Inspired Loss for Enhanced Semantic Change Detection in Remote Sensing}

\author{
Buddhi~Wijenayake$^{\dagger}$,~\IEEEmembership{Student Member,~IEEE}\orcidlink{0009-0001-2624-0251},
Athulya~Ratnayake,~\IEEEmembership{Student Member,~IEEE}\orcidlink{0009-0008-9582-4606},
Praveen~Sumanasekara,~\IEEEmembership{Student Member,~IEEE}\orcidlink{0009-0000-7044-9944},
Roshan~Godaliyadda,~\IEEEmembership{Senior Member,~IEEE}\orcidlink{0000-0002-3495-481X},
Parakrama~Ekanayake,~\IEEEmembership{Senior Member,~IEEE}\orcidlink{0000-0002-5639-8105},
Vijitha~Herath,~\IEEEmembership{Senior Member,~IEEE}\orcidlink{0000-0002-2094-0716},
and~Nichula~Wasalathilaka,~\IEEEmembership{Student Member,~IEEE}\orcidlink{0009-0003-5662-009X}.

\thanks{All the authors of this paper are with the Department of Electrical and Electronic Engineering, University of Peradeniya, Sri Lanka. (Email addresses are listed according to the authors’ order. e-mail: e19445@eng.pdn.ac.lk, e19328@eng.pdn.ac.lk, e19391@eng.pdn.ac.lk, roshangodd@ee.pdn.ac.lk, mpb.ekanayake@ee.pdn.ac.lk, vijitha@ee.pdn.ac.lk, e20425@eng.pdn.ac.lk)}
\thanks{$\dagger$ Corresponding Author, email: e19445@eng.pdn.ac.lk}
}
\markboth{Accepted for publication in IEEE Journal of Selected Topics in Applied Earth Observations and Remote Sensing}{}



\maketitle

\begin{abstract}

Semantic Change Detection (SCD) in remote sensing imagery requires models that integrate extensive spatial context for broad geographic patterns, computational efficiency for large-scale datasets, and sensitivity to class-imbalanced land-cover transitions to detect rare or asymmetric changes. Early SCD approaches relied on Convolutional Neural Networks, which excel in local feature extraction but falter in modeling global spatial context due to limited receptive fields. Transformers mitigate this by capturing long-range dependencies via self-attention, yet their quadratic complexity impairs efficiency on vast remote sensing data. Emerging Mamba architectures, based on state-space models, strike a balance with linear complexity and robust long-range modeling, delivering efficient global context capture and improved performance. In this study, we introduce Mamba-FCS, an SCD framework leveraging a Visual State Space Model backbone, with three key contributions: (1) a Joint Spatio-Frequency Fusion block that integrates log-amplitude frequency-domain features to sharpen edges and mitigate illumination artifacts, (2) a Change-Guided Attention (CGA) module that explicitly bridges the intertwined Binary Change Detection and SCD tasks, and (3) a novel loss function inspired by the Separated Kappa (SeK) metric to optimize for class imbalance. Experiments on the benchmark datasets show that Mamba-FCS consistently outperforms recent state-of-the-art algorithms. Ablation studies indicate that spatio–frequency fusion and CGA mainly sharpen boundaries and suppress hallucinated changes, while the SeK-inspired loss improves minority-class semantics. These results highlight the potential of Mamba-FCS as a scalable and accurate approach for remote sensing change detection. Source code and configuration files are available at \url{https://github.com/Buddhi19/Mamba-FCS.git}

\end{abstract}
\newcommand{\ours}[0]{Mamba-FCS}


\section{Introduction}

Change Detection (CD) is a widely studied and increasingly popular field in remote sensing, which involves identifying alterations in the Earth's surface by comparing remote sensing images of the same area captured at different times\cite{A_Zhu_2024}. CD plays a vital role in various applications, including urban planning, land cover and land use (LCLU) analysis, disaster assessment, ecosystem monitoring, and natural resource management\cite{B_Zhang_2024,R_Wellmann_2020,D_COPPIN_2002,A_Chen_2020}.

CD tasks are primarily categorized into Binary Change Detection (BCD) and Semantic Change Detection (SCD), which are inherently coupled\cite{chengChangeDetectionMethods2024a,tanCGMNetSemanticChange2024}. BCD focuses on detecting whether a change has occurred or not between remote sensing images, classifying areas as either "change" or "no change." In contrast, SCD goes a step further by identifying the specific nature of the change, providing detailed “from-to” transition information between different land cover or land use types\cite{M_Wang_2025}.

Furthermore, CD algorithms can be further divided into supervised and unsupervised approaches. Unsupervised methods such as K-means, ISODATA and graph cut methods aim to identify change labels without the reliance on training with labeled data\cite{N_Lv_2019, A__2022, C_Pérez_2005}. However, this tends to limit the performance especially in complex semantic transitions. Supervised methods, on the contrary, leverage labeled data to learn richer representations of these complex transitions in order to gain superior performance\cite{chengChangeDetectionMethods2024a}.

Early CD algorithms relied on image differencing, change-vector analysis and Bayesian fusion of pre/post semantic maps\cite{A_Bruzzone_2000}.  These multi-stage heuristics accumulate errors and struggle with spectral inconsistencies caused by varying illumination, phenology or sensor viewing angles\cite{C_Hussain_2013}.  

Traditional methods attempted to mitigate these spectral inconsistencies using physical or empirical correction methods. Methods such as C-Correction and Minnaert correction as well as Normalization methods are examples of this\cite{C_Song_2001}. While these approaches can improve robustness of traditional methods induced due to variability, they usually rely on additional data such as elevation models or require parameter tuning\cite{mikeladzeEstimationForestCover2020a}. These shortcomings highlight the need for data-driven alternatives for CD.

Introduction of deep learning architectures opened a new avenue for addressing most of these challenges. Early end‐to‐end designs based on convolutional neural networks (CNNs) overcame many pitfalls of handcrafted pipelines, yet their limited receptive field still hampers the capture of long‐range context in large, heterogeneous scenes\cite{S_Gong_2017,C_Chen_2020,D_Chen_2019}. 

The Attention mechanism, in Transformer Models\cite{A_Beyer_No}
 for incorporating long-range dependencies in sequence modeling, was adopted by Vision Transformers(ViT) for modeling global dependencies in computer vision tasks.
These integrate windowed attentions as well as global attention mechanisms for better modeling of global dependencies. 
These methods and variants have successfully been used in CD Algorithms with competitive results\cite{A_Gedara_2022}.
However, their quadratic complexity makes dense, high-resolution predictions computationally expensive, limiting scalability for large-scale deployment.	

State‐space models (SSMs) such as S4\cite{DBLP:journals/corr/abs-2111-00396} and, more recently, Mamba\cite{gu2024mambalineartimesequencemodeling}
 offer an attractive middle ground- they deliver global sequence modelling with linear complexity and hardware‐friendly implementations.
These models have since been adopted for computer vision tasks \cite{zhu2024visionmambaefficientvisual}
 and later for CD \cite{C_Chen_2024,C_Zhang_2025}
with promising performance.

Despite these advances, real‑world deployment still suffers from class imbalance, isolation of semantic and change detection branches, and neglect of frequency domain cues. In particular, because certain semantic transitions often occupy a small fraction of pixels in certain datasets, conventional models tend to be biased toward more prevalent classes\cite{C_Zheng_2021}. Furthermore, although SCD and BCD are naturally interconnected, existing approaches like~\cite{C_Chen_2024,B_Ding_2022} keep change and semantic decoders isolated, preventing mutual reinforcement, omitting the inherent inter-dependency from the learning process, and resulting in blurred boundaries and inconsistent bi-temporal predictions. Moreover, although Fourier‐domain representations are proven to suppress illumination artifacts and emphasize genuine structural differences in BCD\cite{F_Xing_2025}, existing Mamba frameworks for SCD predominantly focus on spatial features, with limited exploration of frequency-domain fusion benefits.

Furthermore, while the Separated Kappa (SeK) coefficient has become the standard metric for evaluating SCD performance since its introduction on the SECOND benchmark \cite{A_Yang_2022}, its potential as a loss function that support backpropagations for guiding the Deep Learning Model, learnable loss function remains unexplored, representing a clear gap given its ability to leverage the benchmark’s structure for improved learning.

In this work, we introduce \textbf{\ours}, a novel approach that integrates frequency-domain feature fusion and change-guided attention within a Mamba backbone to address limitations of purely spatial processing, and decoder isolation. By incorporating SeK inspired loss term, we emphasis semantic consistency within detected change regions. Extensive experiments on the SECOND \cite{A_Yang_2022} and LandSat-SCD \cite{A_Yuan_2022} benchmarks demonstrate substantial improvements in detecting non-salient changes, enhancing recall for rare transitions, and refining semantic boundaries. In summary, our main contributions includes:

\begin{itemize}
	\item  Proposing Joint Spatio-Frequency Fusion Mechanism, that effectively combines spatial context with frequency-domain information across multiple stages, improving feature representation,
	
	\item Proposing a Change-Guided Attention (CGA) module that iteratively incorporates an intermediate change-probability map into each semantic decoder stage, enabling mutual reinforcement between BCD and SCD branches for higher-fidelity change masks and sharper semantic boundaries, and
	
	\item   Proposing the SeK inspired Loss, leveraging the SeK metric as a loss function to directly optimize for semantic consistency within detected change regions, mitigating the impact of class imbalance.
\end{itemize}

The remainder of this paper is organized as follows. Section \ref{sec:related_work} reviews related work and motivates our contributions. Section \ref{sec:methodology} presents the \ours{} framework, detailing the key components including Joint Spatio-Frequency Fusion block, Change-Guided Attention module, and Separated Kappa loss formulation. Section \ref{sec:exp} describes the experimental methodology and datasets. Section \ref{sec:results} presents comprehensive results and ablation analyses. Finally, Section \ref{sec:conclusion} draws conclusions.

\section{Related Work}\label{sec:related_work}
In this section, we review deep learning-based CD methods for high-resolution optical remote sensing imagery, organized into CNN-based, Transformer-based, State Space Model (SSM) based, and frequency-domain-based approaches, and use this overview to highlight the motivation for our study.

\subsection{CNN-Based Methods}
CNNs have been central to early deep learning approaches for CD because they extract local features effectively. Daudt et al.\ \cite{F_Caye_2018} proposed the first fully convolutional networks (FCNs) for BCD, with three variants: FC-Siam-Conc, FC-EF, and FC-Siam-Diff, which use a Siamese CNN architecture to process bi-temporal inputs. Subsequent work built on this foundation. Fang et al.\ \cite{S_Fang_2022} developed a Siamese architecture with a densely connected CNN to strengthen feature interaction between bi-temporal images. Zhang et al.\ \cite{A_Zhang_2020} introduced multi-level, fine-grained detection through deep supervision of differential features at several CNN stages. Shi et al.\ \cite{A_Shi_2022} further improved discriminative power by incorporating attention. Zhao et al.\ \cite{E_Zhao_2023} studied different fusion strategies for BCD in a dual encoder–decoder architecture that improves detection accuracy. Chen et al.\ \cite{E_Chen_2023} proposed an unsupervised sample generation strategy that swaps image patches so CD models can be trained on single-temporal images.

SCD is more challenging than BCD yet has higher practical value \cite{C_Zheng_2022, M_Caye_2019}. Early CNN-based SCD methods used multitask learning to predict both binary change maps and semantic change categories \cite{M_Caye_2019}. Ding et al.\ \cite{B_Ding_2022} proposed a CNN architecture that increases information exchange between semantic and change decoders and improves SCD performance.

Despite these advances, CNN-based models struggle with long-range context in CD. Their limited receptive fields restrict the capture of large-scale semantic dependencies that are essential for detecting changes across wide regions. This limitation has motivated Transformer-based methods, which provide a global view of the scene \cite{DBLP:journals/corr/VaswaniSPUJGKP17}, and it also motivates architectures that model global context efficiently while enforcing tighter interaction between binary and semantic change branches.

\subsection{Transformer-Based Methods}
The introduction of Vision Transformers (ViT) \cite{A_Beyer_No} for long-range dependency modeling in images has led to growing use of transformer architectures in CD. For BCD, Chen et al.\ \cite{R_Chen_2022} were among the first to apply transformers by converting multi-temporal images into semantic tokens and modeling spatial–temporal relationships to improve detection accuracy. Bandara et al.\ \cite{A_Gedara_2022} then proposed a pure transformer-based Siamese network with a Siamese transformer encoder and an MLP decoder, removing the need for convolutional feature extractors.

For SCD, Ding et al.\ \cite{J_Ding_2024} introduced the semantic change transformer, an adaptation of the CSWin Transformer that explicitly learns semantic transitions. It uses a triple encoder–decoder architecture that enhances spatial features and integrates spatio-temporal dynamics with task-specific priors to reduce learning disparities. SMBCNet \cite{S_Feng_2023} uses a transformer-based architecture for change detection through semantic segmentation, combining a cross-scale enhancement module with a multi-branch change fusion module to capture global information and handle diverse change types. STGNet \cite{M_Wang_2025} further improves SCD by guiding multitask learning through spatio-temporal semantic interaction, refining spatial details and using a bidirectional guidance module to enhance feature extraction in complex scenes.

Transformer-based methods significantly advance CD, yet their quadratic self-attention complexity is problematic for large-scale remote sensing datasets. This has driven interest in alternative architectures such as SSMs \cite{DBLP:journals/corr/abs-2111-00396}.

\begin{figure*}[t]
    \centering
    \includegraphics[width=1\linewidth,height=0.96\textheight,keepaspectratio]{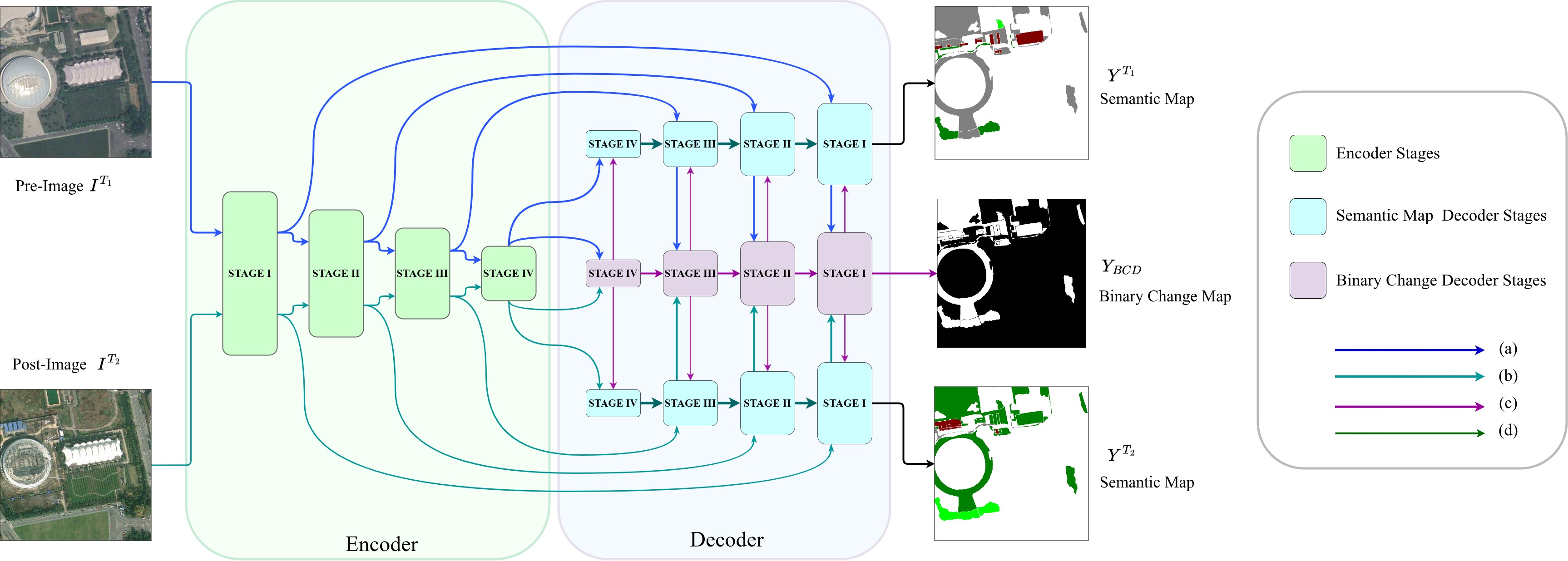}
    \caption{Overview of the proposed \ours{} architecture. A Siamese encoder processes the pre-change image $I^{T_1}$ and post-change image $I^{T_2}$ through four stages (Stage~I–IV) to extract multi-scale features. These features feed a central binary change decoder, which predicts the binary change map $Y_{\mathrm{BCD}}$, and two symmetric semantic decoders that output the semantic maps $Y^{T_1}$ and $Y^{T_2}$. Arrows of type (a) and (b) denote multi-scale skip connections from the pre-change and post-change encoder branches, respectively, to their corresponding decoder stages, (c) indicates within-branch propagation in the semantic decoders, and (d) marks change-guidance connections, where features from the binary decoder are injected into both semantic decoders.}
    
    \label{fig:full_architecture}
\end{figure*}

\subsection{SSM-Based Methods}
SSMs, and especially the Mamba architecture, are promising alternatives to Transformers because they offer linear computational complexity while still modeling long-range dependencies \cite{gu2024mambalineartimesequencemodeling}. In computer vision, Visual State Space Model (VMamba) \cite{liu2024vmambavisualstatespace} and Vision Mamba \cite{zhu2024visionmambaefficientvisual} have enabled SSMs to be used for change detection.

For BCD, MambaBCD \cite{C_Chen_2024} uses VMamba to extract global spatial features from bi-temporal images and employs a change decoder with spatio-temporal relationship modeling that yields strong performance. CDMamba \cite{C_Zhang_2025} introduces the Scaled Residual ConvMamba (SRCM) block, which combines Mamba’s global feature extraction with convolutional layers to strengthen local detail capture, and an Adaptive Global Local Guided Fusion (AGLGF) block that improves bi-temporal feature interaction. MSCNet \cite{M_Sun_} uses a Mamba-based self-correction network with a spatial–channel interaction fusion architecture and a momentum update strategy to generate pseudo-labels and correct noisy annotations, improving performance on challenging BCD tasks. The Iterative Mamba Diffusion Change-Detection Model (IMDCD) \cite{I_Liu_2024} couples Mamba with diffusion models to refine change maps iteratively and increase sensitivity to subtle changes in complex scenes.

For SCD, MambaSCD \cite{C_Chen_2024} adapts VMamba to model complex multi-temporal relationships and effectively capture semantic transitions.

Most existing Mamba-based methods operate only in the spatial domain and rarely exploit frequency-domain information, which could suppress illumination artifacts and emphasize structural differences in multi-temporal imagery.

\subsection{Frequency‐Domain–Based Methods}
Frequency-domain methods transform images from the spatial domain using the Fourier transform and have become a robust option for remote sensing CD. The frequency representation highlights structural differences such as edges and textures and suppresses illumination artifacts that are common in multi-temporal imagery \cite{S_Lu_2004}. These methods are effective for detecting texture changes or periodic patterns and for handling multimodal data from sensors such as optical imagers and Synthetic Aperture Radar (SAR) imagery.

For BCD, MFGFNet \cite{M_Yuan_2023} uses multiple global filters in the frequency domain to enhance boundary sharpness and preserve edges in change regions. By combining frequency-domain processing with multi-scale feature extraction, MFGFNet improves accuracy on datasets such as LEVIR-CD \cite{Chen2020}. FEMCD \cite{F_Xing_2025} introduces a difference-guided state-space model (DGSSM) to extract change-related features and a DCT-aided Mamba decoder (DCTMD) that refines minor and texture changes using frequency cues, achieving strong BCD performance. SpectMamba \cite{S_Dong_2025} further integrates frequency information with the Mamba architecture to handle high-frequency subtle changes and periodic structural changes in multispectral remote sensing images and obtains improved results.

To the best of our knowledge, there is still limited work on integrating frequency and spatial information within the Mamba architecture for SCD. Existing Mamba-based SCD methods often use separate decoders, which can weaken mutual reinforcement between binary and semantic branches. They also do not use the SeK metric as a minimizable loss.

These observations motivate our Mamba-FCS framework. It performs joint spatial–frequency fusion within a Mamba encoder–decoder, introduces change-guided semantic decoders that share information with the binary branch, and employs a SeK-inspired loss to better capture rare semantic transitions. Unlike existing Transformer and Mamba-based SCD models that mainly use spatial features, keep binary and semantic decoders loosely coupled, and rely on generic losses for class imbalance, Mamba-FCS is designed to exploit frequency-domain cues, enforce semantic consistency through change-guided interaction from the binary change map, and directly optimize a SeK-inspired objective under severe class imbalance.

\begin{figure*}[t]
    \centering
    \includegraphics[width=0.9\linewidth]{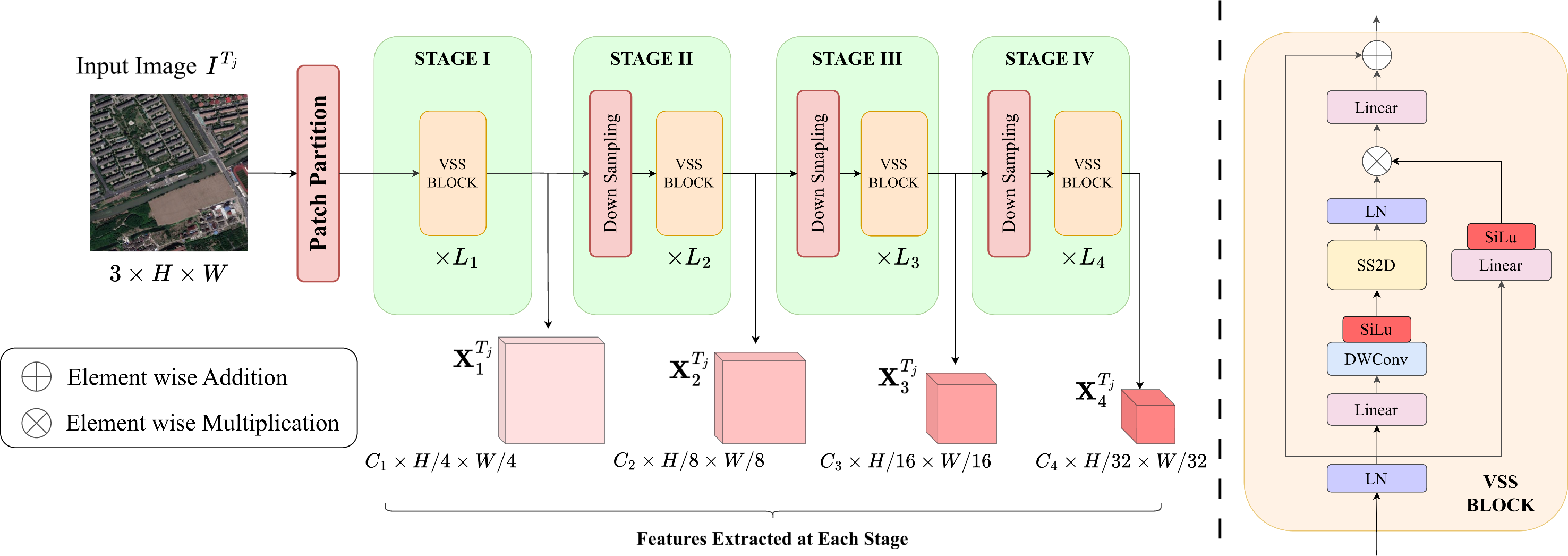}
    \caption{Architecture of the Visual State-Space Model (VMamba) backbone, serving as the shared encoder. Following an initial patch-partition layer, the encoder comprises four stages \(i=1,2,3,4\), each with \(L_i\) Visual State-Space (VSS) blocks. These blocks utilize 2D selective scanning to progressively downsample the spatial resolution (from \(H \times W\) to \(H/32 \times W/32\)) while expanding the channel depth. Feature dimensions extracted from each stage are annotated adjacent to the corresponding blocks.}
    \label{fig:encoder_architecture}
\end{figure*}
\section{Methodology}\label{sec:methodology}
In this section, we first present an overview of the proposed architecture. We then describe each component of the architecture in detail. Finally, we discuss the loss functions employed in our methodology.

\subsection{Overview of the Network Architecture}
The proposed \ours{} framework utilizes a Siamese network architecture with a shared encoder for processing bi-temporal imagery, followed by three decoders to jointly perform SCD and BCD. As illustrated in Figure \ref{fig:full_architecture}, the framework processes input image pairs $I^{T_1}, I^{T_2} \in \mathbb{R}^{3 \times H \times W}$, where 3 denotes the RGB color channels, and $H$ and $W$ represent the image height and width respectively, corresponding to pre-change and post-change temporal scenes. These bi-temporal inputs are processed by a shared Visual State Space Model (VMamba) based backbone encoder, denoted as $\mathcal{F}_{encoder}$, which is detailed in Section~\ref{sec:encoder}. This encoder, as seen in figure \ref{fig:encoder_architecture} extracts multi-scale feature maps at four hierarchical levels, providing a robust foundation for subsequent decoding and fusion operations. The feature extraction process is formalized as follows.

\begin{equation}
    \begin{aligned}
    &X_1^{T_1}, X_2^{T_1}, X_3^{T_1}, X_4^{T_1} = \mathcal{F}_{encoder}(I^{T_1}) \\
    &X_1^{T_2}, X_2^{T_2}, X_3^{T_2}, X_4^{T_2} = \mathcal{F}_{encoder}(I^{T_2})
    \end{aligned}
    \label{eq:encoder}
\end{equation}

where $X_i^{T_1}, X_i^{T_2} \in \mathbb{R}^{C_i \times H_i \times W_i}$ denote the feature maps extracted from the $i$-th stage of the encoder for time steps $T_1$ and $T_2$, respectively, with $C_i$, $H_i$, and $W_i$ representing the channel depth, height, and width at each stage.

These multi-scale feature maps are then fed to the three decoder networks as seen in figure \ref{fig:full_architecture}. First, our Binary Change Decoder, $\mathcal{F}_{BCD}$, detailed in section \ref{sec:bcd}, detects changes between the two time steps by fusing corresponding feature maps from $I^{T_1}$ and $I^{T_2}$ at each stage. The fusion is performed using our novel Joint Spatio-Frequency Feature Fusion Mechanism $\mathcal{F}_{fusion}$ as described in section \ref{sec:fusion}. As illustrated in figure \ref{fig:bcd_decoder}, the Binary Change Decoder employs a hierarchical refinement strategy to generate the final binary change detection output $Y_{BCD} \in \mathbb{R}^{2 \times H \times W}$, where the two channels correspond to change and no-change classes. Additionally, the decoder produces intermediate change maps $CM_i \in \mathbb{R}^{C_i \times H_i \times W_i}$ for $i \in \{1,2,3,4\}$ representing each hierarchical level. This can be formulated as,

\begin{equation}
    Y_{\text{BCD}}, \{CM_i\}_{i=1}^4= \mathcal{F}_{BCD}(\{X_i^{T_1}, X_i^{T_2}\}_{i=1}^4)
    \label{eq:bcd}
\end{equation}
The change maps $\{CM_i\}_{i=1}^4$ will be used in our CGA Module in Semantic Change Detection, detailed in the section \ref{sec:cga}.

For SCD, the framework incorporates two dedicated Semantic Map Decoders, $\mathcal{F}_{SCD}^{T_1}$ and $\mathcal{F}_{SCD}^{T_2}$, one for each time step as detailed in section \ref{sec:smd}. These decoders, as illustrated in figure \ref{fig:scd_decoder}, utilize encoder extracted features $\{X_i^{T_j}\}_{i=1}^4$ ($j \in \{1, 2\}$) and intermediate change maps $\{CM_i\}_{i=1}^4$ extracted from Binary Change Detector to enhance semantic segmentation by prioritizing changed regions (The information flow can be seen in figure \ref{fig:full_architecture}). The process can be defined as,

\begin{equation}
    \begin{aligned}
    &Y^{T_1} = \mathcal{F}_{SCD}^{T_1}(\{X_i^{T_1}\}_{i=1}^4, \{CM_i\}_{i=1}^4) \\
    &Y^{T_2} = \mathcal{F}_{SCD}^{T_2}(\{X_i^{T_2}\}_{i=1}^4, \{CM_i\}_{i=1}^4)
    \end{aligned}
    \label{eq:scd}
\end{equation}
where $Y^{T_1}, Y^{T_2}\in\mathbb{R}^{N\times H\times W}$ are the semantic segmentation maps at $T_1$ and $T_2$, respectively, with \(N\) channels corresponding to \(N\) semantic classes.

\begin{figure*}[t]
    \centering
    \includegraphics[width=\linewidth]{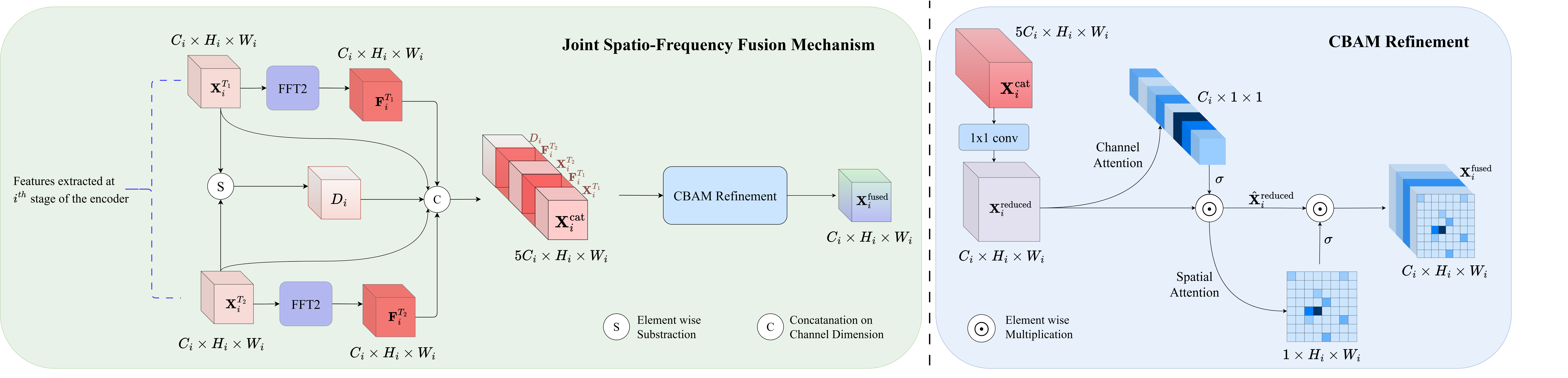}
    \caption{(a) The Joint Spatio-Frequency Feature Fusion (\(F_{\text{fusion}}\)) block, embedded within each decoder stage $i$, concatenates spatial features \(X^{T_1}_i\) and \(X^{T_2}_i\), log-amplitude frequency-domain features \(F^{T_1}_i\) and \(F^{T_2}_i\), and the absolute difference map \(D_i\) to form $X_i^{cat}$. (b) The CBAM Refinement Module compresses and refines $X_i^{cat}$ through a Convolutional Block Attention Module (CBAM), yielding the fused output tensor \(X^{\text{fused}}_i\).}
    \label{fig:fusion}
\end{figure*}
\subsection{Visual State Space Model Based Backbone Encoder- VMamba}\label{sec:encoder}
The backbone encoder in the proposed \ours{} framework is based on Visual State Space Models (VMamba) \cite{liu2024vmambavisualstatespace}, a state-of-the-art (SOTA) vision backbone designed for efficient visual representation learning with linear computational complexity. VMamba adapts the Mamba state-space model, originally developed for sequence modeling in natural language processing, to the vision domain, making it particularly suitable for processing bi-temporal imagery due to its ability to capture long-range dependencies with minimal computational overhead \cite{gu2024mambalineartimesequencemodeling}. Its promising adaptation to CD tasks is well described in \cite{C_Chen_2024}.

As depicted in Figure \ref{fig:encoder_architecture}, the VMamba encoder employs a hierarchical architecture to process input images. The process begins with a Patch Partition module that divides the input image $ I^{T_j}$, ($j\in\{1,2\}$) into patches while embedding them into a higher-dimensional feature space. This is followed by four hierarchical stages, each consisting of $ L_i $ Visual State-Space (VSS) blocks and, except for the first stage, a down-sampling layer. These stages generate feature maps at progressively reduced resolutions ($H/4\times W/4$, $ H/8 \times W/8 $, $ H/16 \times W/16 $, $ H/32 \times W/32 $) with increasing channel depths ($ C_1< C_2 < C_3 <C_4 $), facilitating the extraction of multi-scale features essential for both BCD and SCD.

As seen in the right panel of Figure \ref{fig:encoder_architecture}, each VSS block incorporates a 2D Selective Scan (SS2D)
module, which replaces the traditional self-attention mechanism found in vision transformers. The SS2D module employs
a Cross-Scan strategy, scanning the 2D feature map along
four directions (top-left to bottom-right, bottom-right to top-left, top-right to bottom-left, and bottom-left to top-right). This approach bridges the gap between the
sequential nature of 1D state-space models and the nonsequential structure of 2D visual data, ensuring a global
receptive field while maintaining linear complexity\cite{liu2024vmambavisualstatespace}.

VMamba has 3 variants named VMamba-Small, VMamba-Tiny and VMamba-Base. In our implementation, we utilize the VMamba-Base variant, which outperforms other variants with configuration given in table \ref{tab:vmamba_base_config}.

\begin{table}[h]
	\centering
	\caption{Configuration details for stages $i \in \{1, 2, 3, 4\}$. $C_i$ denotes the channel depth, $L_i$ represents the number of VSS blocks in each encoder stage, and $H_i$ and $W_i$ denote the height and width relative to the original image dimensions $H$ and $W$.}
	\label{tab:vmamba_base_config}
	\begin{tabular}{c|cccc}
		\toprule
		Stage - \(i\) & 1 & 2 & 3 & 4 \\
		\midrule
		\(C_i\) & 128 & 256 & 512 & 1024 \\
		\(L_i\) & 2 & 2 & 15 & 2 \\
		\(H_i\) & \(H/4\) & \(H/8\) & \(H/16\) & \(H/32\) \\
		\(W_i\) & \(W/4\) & \(W/8\) & \(W/16\) & \(W/32\) \\
		\bottomrule
	\end{tabular}
\end{table}

\subsection{Joint Spatio-Frequency Feature Fusion Mechanism}
\begin{figure*}[t]
    \centering
    \includegraphics[width=\linewidth]{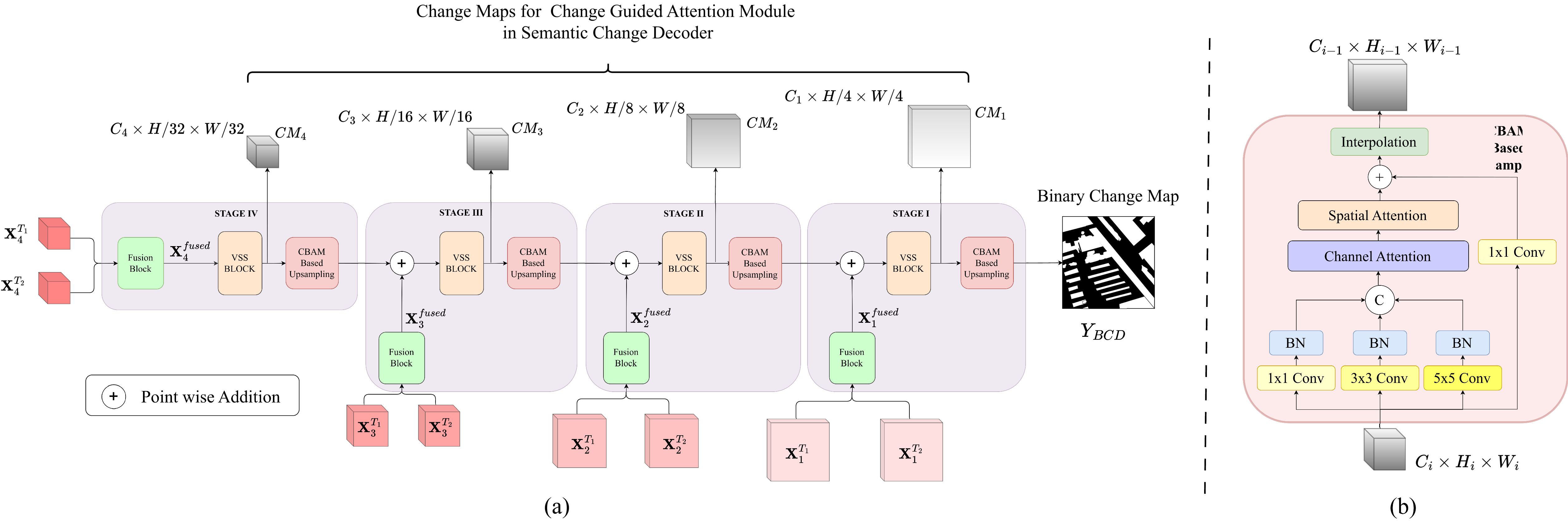}
    \caption{(a) Architecture of the Binary Change Decoder for generating the Binary Change Map $Y_{BCD}$. At each stage, encoder features from two time points ($X_i^{T_1}$ and $X_i^{T_2}$) are fused through a fusion block to obtain $X^{\text{fused}}_i$. The fused features are then passed through a VSS block followed by a CBAM-based upsampling unit. Point-wise addition progressively integrates multi-scale information, while intermediate change maps $\{CM_i\}_{i=1}^4$ are extracted to support the Change-Guided Attention (CGA) module. (b) The architecture of the CBAM-based Upsampling Block, which reduces $C_i$ and increases the $H_i$ and $W_i$ for the next stage.}
    \label{fig:bcd_decoder}
\end{figure*}

{\label{sec:fusion}}
As illustrated in Figure~\ref{fig:fusion}, we propose a novel fusion mechanism $\mathcal{F}_{fusion}$ to enhance CD in bi-temporal images that jointly exploits spatial cues and frequency cues to highlight subtle scene variations. At each stage $i$ in the Binary Change Decoder, the feature maps from pre-change ($X_i^{T_1}$) and post-change ($X_i^{T_2}$) images are fused using the Fusion Block as illustrated in figure \ref{fig:bcd_decoder} to form $X_i^{\mathrm{fused}}\in\mathbb{R}^{C_i\times H_i\times W_i}$ as follows,

\subsubsection{FFT2 Branch}
To capture high-frequency components, emphasizing subtle changes like edges and textures, we incorporate the FFT2 Branch, that transforms the spatial information into the frequency domain as,

\begin{equation}
	\begin{aligned}
		F_i^{T_1} &= \log(1 + |\text{FFT2}(X_i^{T_1})|)\\
		F_i^{T_2} &= \log(1 + |\text{FFT2}(X_i^{T_2})|)
	\end{aligned}
\end{equation}

where \(\mathrm{FFT2}\)\footnote{Implemented with \texttt{torch.fft.fft2} in PyTorch, using \texttt{norm='ortho'} for orthonormal scaling.} denotes the 2-dimensional (2D) Fast Fourier Transform (FFT)~\cite{fft}
, which operates channel-wise.  
For a single channel \(k \in \{1, 2, \dots, C_i\}\), \(X_{i,k}^{T_j} \in \mathbb{R}^{H_i \times W_i}\), the 2D FFT is defined as,

\begin{equation}
	\footnotesize
	\mathcal{F}\!\bigl(X_{i,k}^{T_j}\bigr)(u,v) \;=\;
	\frac{1}{\sqrt{H_i W_i}}
	\sum_{m=0}^{H_i-1} \sum_{n=0}^{W_i-1}
	X_{i,k}^{T_j}(m,n)\;
	e^{-\,2\pi i \!\left(\frac{u m}{H_i} + \frac{v n}{W_i}\right)}
	\label{eq:unitary_dft}
\end{equation}

Here, \(m\) and \(n\) are spatial indices, and \(u = 0, 1, \dots, H_i-1\) and \(v = 0, 1, \dots, W_i-1\) are the frequency indices corresponding to the spatial dimensions \(H_i\) and \(W_i\), respectively. The absolute value of the FFT2 is taken, and a logarithmic scaling is applied to emphasize high-frequency components while compressing the dynamic range, making the representation more robust for detecting subtle changes.

\subsubsection{Difference Branch}
We compute the spatial difference between the pre- and post-event feature maps to highlight potential change regions as follows,

\begin{equation}
D_i = |X_i^{T_1} - X_i^{T_2}|
\end{equation}

\subsubsection{Attention-Based Spatial-Frequency Fusion}

The spatial features $X_i^{T_1}$, $X_i^{T_2}$, frequency features $F_i^{T_1}$, $F_i^{T_2}$, and difference feature $D_i$ are concatenated along the channel axis as follows:
\begin{equation}
	X_i^{\mathrm{cat}} = \text{Concat}\bigl(X_i^{T_1}, F_i^{T_1}, X_i^{T_2}, F_i^{T_2}, D_i\bigr).
\end{equation}

The concatenated feature $X_i^{\mathrm{cat}}$ is compressed using a $1\times1$ convolution, yielding a reduced feature map $X_i^{\mathrm{reduced}} \in \mathbb{R}^{C_i \times H_i \times W_i}$.

To enhance informative features and suppress noise, we refine the reduced feature map $X_i^{\mathrm{reduced}}$ using the Convolutional Block Attention Module (CBAM)~\cite{C_Woo_2018}, as illustrated in Fig.~\ref{fig:fusion}\textcolor{blue}{(b)}. CBAM first applies channel attention to produce $\mathbf{g}^C_i$ as,
\begin{equation}
	\mathbf{g}^C_i = \sigma\left(\mathrm{MLP}\big(\mathrm{GAP}(X_i^{\mathrm{reduced}})\big) + \mathrm{MLP}\big(\mathrm{GMP}(X_i^{\mathrm{reduced}})\big)\right),
	\label{eq:cbam-channel}
\end{equation}
where $\sigma$ is the sigmoid activation, $\mathrm{GAP}$ and $\mathrm{GMP}$ are global average and max pooling over spatial dimensions, and the two pooled vectors are passed through a shared bottleneck MLP~\cite{C_Woo_2018}. The resulting channel gate is applied to $X_i^{\mathrm{reduced}}$ as
\begin{equation}
	\hat{X}_i^{\mathrm{reduced}} = \mathbf{g}^C_i \odot X_i^{\mathrm{reduced}},
	\label{eq:cbam-channel-applied}
\end{equation}
with $\odot$ denoting element-wise multiplication.

Next, spatial attention is computed on $\hat{X}_i^{\mathrm{reduced}}$:
\begin{equation}
	\mathbf{g}^S_i = \sigma\left(f^{7\times 7}\big(\mathrm{Avg}_c(\hat{X}_i^{\mathrm{reduced}}) \textcircled{c} \mathrm{Max}_c(\hat{X}_i^{\mathrm{reduced}})\big)\right),
	\label{eq:cbam-spatial}
\end{equation}
where $\mathrm{Avg}_c$ and $\mathrm{Max}_c$ perform average and max pooling along the channel dimension, $f^{7\times7}$ is a $7\times7$ convolution, and $\textcircled{c}$ denotes channel-wise concatenation. The final fused representation is
\begin{equation}
	X_i^{\mathrm{fused}} = \mathbf{g}^S_i \odot \hat{X}_i^{\mathrm{reduced}}.
	\label{eq:cbam-output}
\end{equation}
Intuitively, $\mathbf{g}^C_i$ reweights feature channels based on global context (blue bars in Fig.~\ref{fig:fusion}\textcolor{blue}{(b)}), while $\mathbf{g}^S_i$ emphasizes salient spatial locations (blue squares), yielding the refined fused representation $X_i^{\mathrm{fused}}$.

\begin{figure*}[t]
    \centering
    \includegraphics[width=\linewidth]{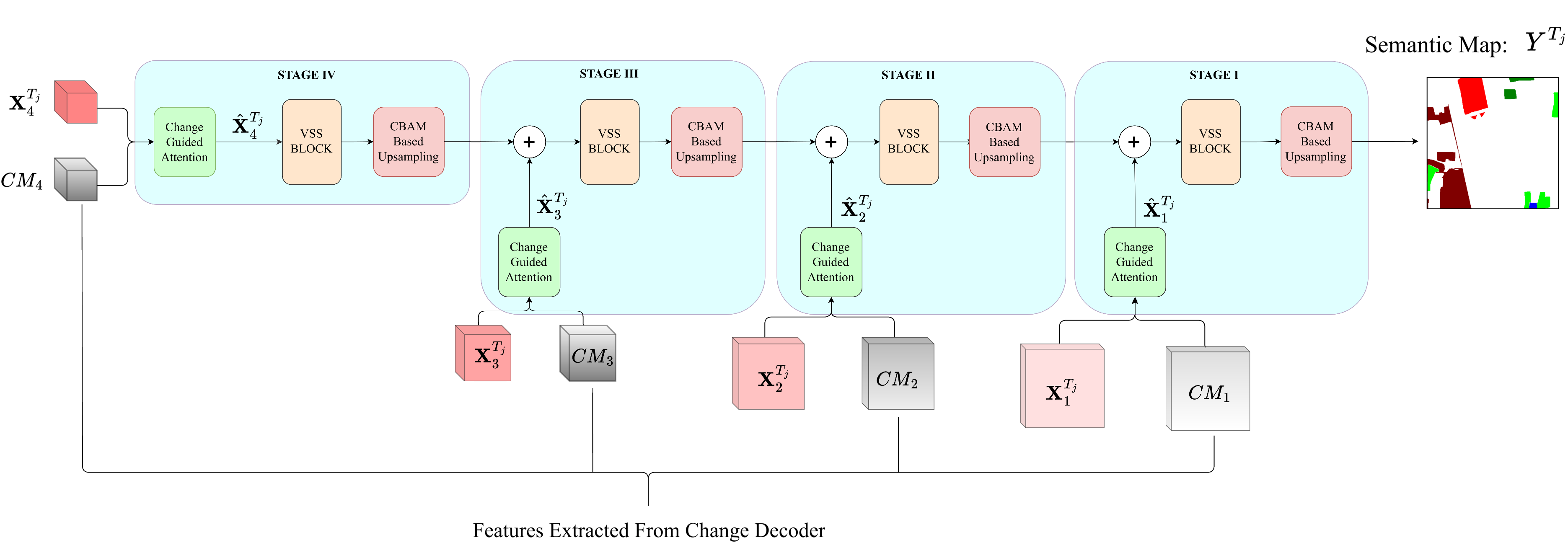}
    \caption{(a) Architecture of the Semantic Map Decoder for the $j^{\text{th}}$ time stamp ($j \in \{1,2\}$). At each stage, the encoder feature $X_i^{T_j}$ is refined using the corresponding change map $CM_i$ through a Change-Guided Attention module, producing $\hat{X}_i^{T_j}$. Point-wise addition is employed at Stages I--III to progressively integrate multi-scale information. The refined features are then processed by a VSS block and upsampled via a CBAM-based upsampling unit. The final output is the semantic map $Y^{T_j}$ for timestamp $T_j$.}
    \label{fig:scd_decoder}
\end{figure*}
\subsection{Binary Change Decoder}\label{sec:bcd}

As seen in figure \ref{fig:bcd_decoder}, the Binary Change Decoder employs a top-down refinement strategy. At the coarsest stage $i{=}4$ we first fuse $X_4^{T_1}$ and $X_4^{T_2}$ using the mechanism of Section~\ref{sec:fusion}, producing $X_4^{\mathrm{fused}}$. A VSS Block then models long-range dependencies within this tensor. The VSS output feeds a CBAM-based upsampling unit.

\paragraph*{CBAM-based Upsampling}\label{CBAM}
As illustrated in Figure~\ref{fig:bcd_decoder}\textcolor{blue}{(b)}, the CBAM-based upsampling unit employs three parallel convolutions with kernels $1{\times}1$, $3{\times}3$, and $5{\times}5$ to gather context at complementary spatial scales, a strategy shown to enlarge the effective receptive field and sharpen object boundaries during decoder up-sampling\cite{chenChangeDetectionMultitemporal2020,ratnayakeEnhancedSCanNetCBAM2025,wijenayakePrecisionSpatioTemporalFeature2025,yinCNNTransformerNetworkCombining2023}. Each branch output is batch-normalized, concatenated, and re-weighted using CBAM’s channel and spatial attention mechanisms as described in equations \eqref{eq:cbam-channel} -- \eqref{eq:cbam-output}, which adaptively suppresses noise. The result is combined with a convolutional branch of kernel size $1\times1$ and interpolated to align with the H, W, and C of the next stage, before being passed to the subsequent, finer stage.

For each finer scale $i\in\{3,2,1\}$, we compute $X_i^{\mathrm{fused}}$ by fusing the corresponding encoder features as illustrated in \ref{fig:fusion}. This fused representation is added element-wise to the upsampled output from the previous stage. The sum is then fed into a VSS block to model multi-scale context, and its output is passed through the CBAM-based upsampling module to generate features for the next stage. 
Additionally, we extract the VSS block output at each scale as $CM_i$, which serves as input to the CGA module.

\subsection{Semantic Map Decoders}\label{sec:smd}
The Semantic Map Decoders follow a similar top-down, multi-scale refinement strategy as $\mathcal{F}_{\text{BCD}}$, to produce timestamp-specific semantic maps that are \emph{conditioned} on the change cues generated by $\mathcal{F}_{\text{BCD}}$ as seen in figure \ref{fig:scd_decoder}. Two weight-independent decoders, $\mathcal{F}_{\text{SCD}}^{T_1}$ and $\mathcal{F}_{\text{SCD}}^{T_2}$, share an identical architecture, allowing each to learn class priors specialised to the pre and post-change scenes, respectively (see Eq.\ref{eq:scd}).

\subsubsection{\textbf{Change-Guided Attention (CGA)}}\label{sec:cga}
At each stage $i$, the raw encoder feature $X_i^{T_j}$, ($j\in \{1,2\}$) is first modulated by its corresponding auxiliary change map $CM_i$ using the Change-Guided Attention block as follows,

\begin{equation}
\widehat{X}_i^{T_j} = X_i^{T_j} \;\odot\; \sigma(CM_i)
\end{equation}

where $\sigma$ denotes the sigmoid function and $\odot$ indicates element-wise multiplication. This mechanism guides $\mathcal{F}_{\text{SCD}}$ to focus on regions where changes are likely to occur.

At stage $i=4$, we compute $\widehat{X}_4^{T_j}$ and pass it through a VSS Block. The output of the VSS Block is then fed into a CBAM-based upsampling block, preparing it for the next stage.

At each finer stage $i \in \{3, 2, 1\}$, we compute $\widehat{X}_i^{T_j}$ and perform element-wise addition with the upsampled output from the previous stage. The resulting sum is then passed through a VSS Block, followed by a CBAM-based upsampling block(see section \ref{CBAM}), to continue the decoding process.

\subsection{Loss Function}\label{sec:loss}

Our network is trained using a composite loss that integrates widely adopted cross-entropy loss\cite{mao2023crossentropylossfunctionstheoretical}
 and a mean Intersection over Union (\(\mathrm{mIoU}\)) regulariser\cite{O_Atiqur_2016}
, alongside our novel Separated Kappa (SeK)\cite{A_Yang_2022} inspired loss, which explicitly rewards semantic consistency within detected change regions.

For each output \(b \in \{Y_{\text{BCD}}, Y^{T_1}, Y^{T_2}\}\) we minimise the pixel-wise cross-entropy loss,
\begin{equation}
	\mathcal{L}_{\mathrm{CE}}^{b} = -\frac{1}{N_b}
	\sum_{i=1}^{N_b}\sum_{c=0}^{C_b-1}
	\tilde{Y}^{b}_{i}(c)\,
	\log\!\bigl(P^{b}_{i}(c)\bigr),
	\label{eq:ce_generic}
\end{equation}
where \(N_b\) is the number of valid (non-void) pixels, \(C_b=2\) for the binary change map and \(C_b\) equals the number of land-cover classes for the semantic maps, $ P^{b}_{i}(c) $ is the softmax probability that pixel $ i $ belongs to class $ c $ in branch $ b $, $ \tilde{Y}^{b}_{i}(c) $ is the one-hot encoded ground-truth tensor, equal to 1 if pixel $ i $ belongs to class $ c $, and 0 otherwise.

To address the inherent class imbalance in change detection tasks, where unchanged pixels typically dominate the dataset, and to improve boundary delineation accuracy, we incorporate an mIoU regularizer \cite{O_Atiqur_2016}. The mIoU metric provides balanced evaluation across all classes by computing the intersection over union for each class independently, thereby preventing the model from being biased toward the majority class. This regularization term has been shown to significantly improve segmentation performance\cite{zhong2024lrnetchangedetectionhighresolution}.

\begin{figure*}[t]
	\centering
	\includegraphics[width=1\linewidth]{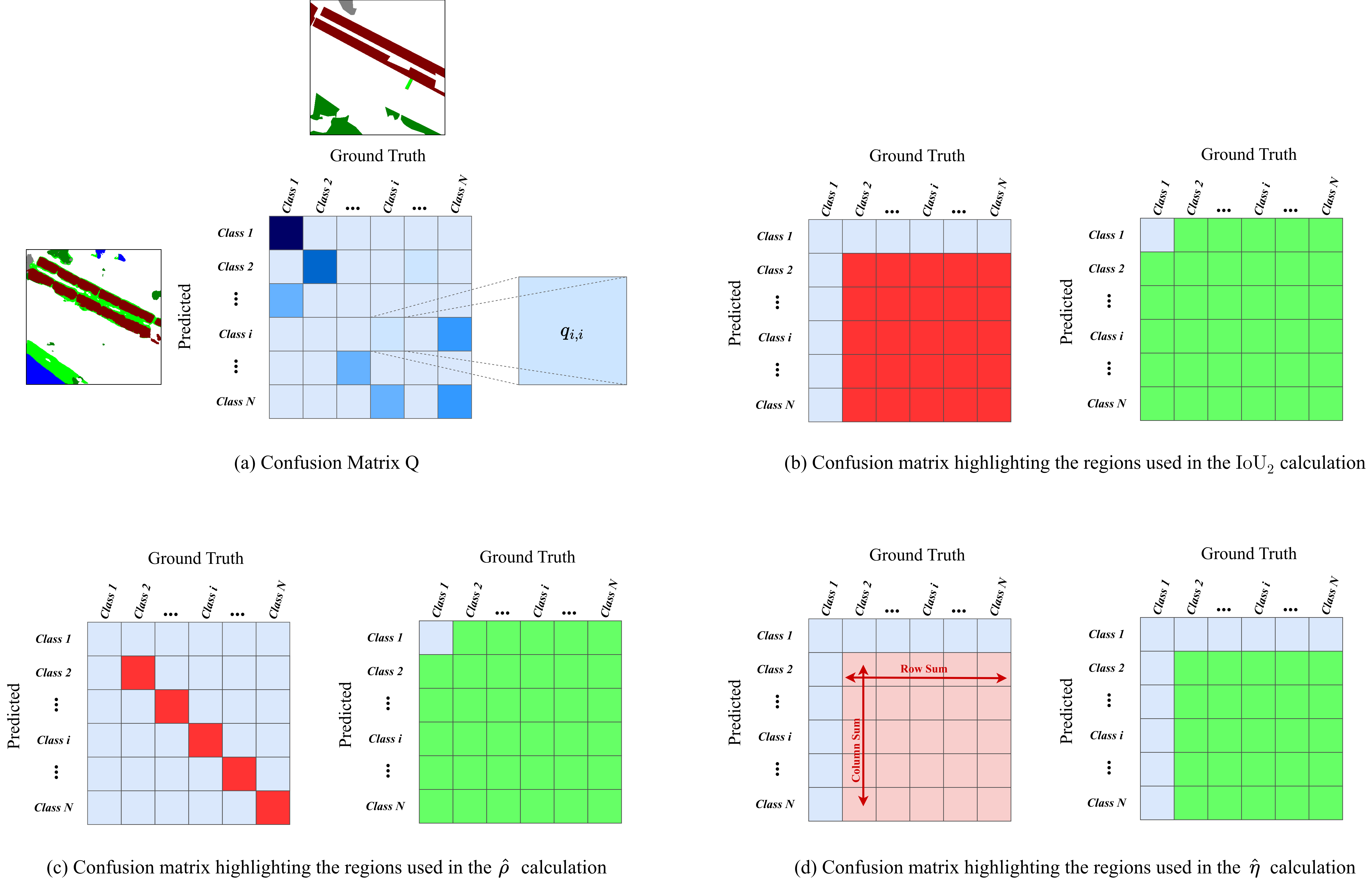}
	\caption{Illustration of the confusion matrix $Q$ and the regions used in computation of SeK. 
		(a) Full confusion matrix $Q$ with entries $q_{i,j}$. 
		(b) $\mathrm{IoU}_2$ -- the sum of \textcolor{red}{red cells} represent the numerator (all predicted–change vs.\ true–change entries), and the sum of \textcolor{green}{green cells} represent the denominator (all entries excluding $q_{11}$). 
		(c) $\hat{\rho}$ -- the sum of \textcolor{red}{red diagonal cells} correspond to the numerator (correctly classified change classes), and the sum of \textcolor{green}{green cells} correspond to the denominator (all change-related entries, i.e.\ all except $q_{11}$). 
		(d) $\hat{\eta}$ -- the numerator is computed from sum of the products of row and column sums of the \textcolor{red}{red region}, while the square of the sum of \textcolor{green}{green cells} indicates the denominator.}
	\label{fig:sek_confusion}
\end{figure*}

As illustrated in figure \ref{fig:sek_confusion}\textcolor{blue}{(a)}, $Q \in \mathbb{R}^{N \times N}$ represent the confusion matrix derived from predicted and Ground Truth (GT) labels, where $q_{i,j}$ denotes the count of pixels classified as class $i$ by the model but belonging to GT class $j$, for $i,j\in \{1,\cdots,N\}$,with class 1
signifying no-change. We define mIoU loss as,
\begin{equation}
	\mathcal{L}_{\text{mIoU}} = -\log \bigl( \mathrm{mIoU} + \varepsilon \bigr),
	\label{eq:miou_loss}
\end{equation}
where,
\begin{equation}
	\mathrm{IoU}_{1} = \frac{q_{11}}{\sum_{k=1}^{N} q_{k1} + \sum_{k=1}^{N} q_{1k} - q_{11}},
	\label{eq:iou1}
\end{equation}
\begin{equation}
	\mathrm{IoU}_{2} = \frac{\sum_{i=2}^N\sum_{j=2}^Nq_{ij}}{\sum_{i=1}^{N} \sum_{j=1}^{N} q_{ij} - q_{11}},
	\label{eq:iou2}
\end{equation}
\begin{equation}
	\mathrm{mIoU} = \frac{1}{2} \bigl( \mathrm{IoU}_{1} + \mathrm{IoU}_{2} \bigr),
\end{equation}
with \(\varepsilon = 10^{-6}\) ensuring numerical stability.

\subsubsection{Seperated Kappa}\label{sec:SeK}
The \emph{Separated Kappa} (SeK) metric, an adaptation of Cohen’s $\kappa$ tailored for semantic change detection, quantifies semantic agreement between predicted and ground-truth class labels solely within regions marked as changed, effectively excluding unchanged pixels\cite{A_Yang_2022}. The SeK metric is formulated as,
\begin{equation}
\mathrm{SeK}
   = \exp\!\bigl(\mathrm{IoU}_2 - 1\bigr)\,
     \frac{\hat{\rho}-\hat{\eta}}{1-\hat{\eta}}
     \label{sek-eq}
\end{equation}
where,

\begin{equation}
\hat{\rho} = \frac{\displaystyle\sum_{i=2}^{N}q_{ii}}
                     {\displaystyle\sum_{i=1}^{N}\sum_{j=1}^{N}q_{ij} - q_{11}}
\end{equation}

\begin{equation}
	\hat{\eta} \;=\;
	\frac{\displaystyle \sum_{j=2}^{N}
		\Big(\sum_{i=2}^{N} q_{j i}\Big)\Big(\sum_{i=2}^{N} q_{i j}\Big)}
	{\Big(\displaystyle \sum_{i=2}^{N}\sum_{j=2}^{N} q_{i j}\Big)^2} .
\end{equation}

As illustrated in figure ~\ref{fig:sek_confusion}\textcolor{blue}{(b)}, $\mathrm{IoU}_2$ quantifies the overlap between predicted and true change pixels. In figure ~\ref{fig:sek_confusion}\textcolor{blue}{(c)}, $\hat{\rho}$ measures semantic accuracy within change regions by focusing on the diagonal elements of the reduced confusion matrix. As shown in figure \ref{fig:sek_confusion}\textcolor{blue}{(d)}, $\hat{\eta}$ accounts for chance agreement within the same change-only matrix. Thus, SeK effectively emphasizes semantic consistency in change regions while mitigating the influence of the non-change class.

\subsubsection{\textbf{SeK Inspired Loss}}
We compute the SeK metric at each of two timestamps, $T_1$ and $T_2$, and calculate their average as follows,
\begin{equation}
	\mathrm{SeK}_{\text{avg}} = \frac{1}{2} \left( \mathrm{SeK}_{1} + \mathrm{SeK}_{2} \right).
	\label{eq:sek_avg}
\end{equation}
While SeK can take negative values when agreement is worse than chance, we consistently observed $\mathrm{SeK}_{\text{avg}} > 0$ in our experiments. Nevertheless, to handle rare degenerate cases where $\mathrm{SeK}_{\text{avg}} < 0$ (indicating worse-than-chance agreement that could lead to undefined logarithmic losses and unstable gradients during optimization), we apply a clipping operation as,
\begin{equation}
	\overline{\mathrm{SeK}} = \max \left( \mathrm{SeK}_{\text{avg}}, 0 \right).
	\label{eq:sek_clip}
\end{equation}
The SeK-based loss is then defined as,
\begin{equation}
	\mathcal{L}_{\mathrm{SeK}} = -\log \left( \overline{\mathrm{SeK}} + \varepsilon \right),
	\label{eq:sek_loss}
\end{equation}
where \(\varepsilon=10^{-6}\) ensures numerical stability.

\paragraph*{\textbf{Validity of SeK Inspired Loss}}
For training, we adopt a differentiable “soft confusion matrix’’ formulation of SeK. Let $\Omega$ denote the set of pixels and let $y_x \in \{1,\dots,N\}$ be the ground-truth change type (with $1$ denoting no-change) at pixel $x \in \Omega$. For each pixel, the network outputs logits $z(x)$ and softmax probabilities $p_i(x) = \mathrm{softmax}(z(x))_i$. The $(i,j)$-th entry of the soft confusion matrix $Q$ is then
\begin{equation}
	q_{ij} = \sum_{x \in \Omega} \mathbf{1}[y_x = j]\, p_i(x),
\end{equation}
where $\mathbf{1}[\cdot]$ is the indicator function. With this construction, all quantities entering the SeK expression (the $q_{ij}$, $\mathrm{IoU}_2$, $\hat{\rho}$, $\hat{\eta}$, $\mathrm{SeK}_1$, $\mathrm{SeK}_2$, and $\mathrm{SeK}_{\text{avg}}$) become compositions of sums, products, and ratios of softmax probabilities, and are therefore differentiable with respect to the network logits.

The final loss $\mathcal{L}_{\mathrm{SeK}}$ in \eqref{eq:sek_loss} is obtained by applying a clipping operator and a negative logarithm to $\mathrm{SeK}_{\text{avg}}$. For $\mathrm{SeK}_{\text{avg}} > 0$, we have $\overline{\mathrm{SeK}} = \mathrm{SeK}_{\text{avg}}$ and
\[
\frac{\partial \mathcal{L}_{\mathrm{SeK}}}{\partial \mathrm{SeK}_{\text{avg}}}
= -\frac{1}{\mathrm{SeK}_{\text{avg}} + \varepsilon},
\]
which is finite for all $\mathrm{SeK}_{\text{avg}} > 0$. The clipping at zero introduces a non-differentiable point at $\mathrm{SeK}_{\text{avg}} = 0$, but this is a standard subgradient situation (analogous to ReLU), and in practice we observe $\mathrm{SeK}_{\text{avg}} > 0$ during training. Since $\overline{\mathrm{SeK}} \in [0,1]$, the loss $\mathcal{L}_{\mathrm{SeK}} = -\log(\overline{\mathrm{SeK}} + \varepsilon)$ is bounded between two finite constants and decreases monotonically as semantic agreement in changed regions improves. Hence, the proposed SeK loss is well-defined, bounded from below, and (almost everywhere) differentiable with respect to the model parameters, and can be safely optimized via backpropagation.

\subsubsection{Overall Objective}
The total training objective of our network is defined as,
\begin{equation}
	\begin{aligned}
		\mathcal{L}_{\text{total}} = \mathcal{L}_{\text{CE}}^{Y_{\text{BCD}}} + \frac{1}{2} \left( \mathcal{L}_{\text{CE}}^{Y^{T_1}} + \mathcal{L}_{\text{CE}}^{Y^{T_2}} \right) + \lambda_1 \mathcal{L}_{\mathrm{mIoU}} + \lambda_2 \mathcal{L}_{\mathrm{SeK}},
	\end{aligned}
\end{equation}
where $\mathcal{L}_{\text{CE}}^{Y_{\text{BCD}}}$, $\mathcal{L}_{\text{CE}}^{Y^{T_1}}$, and $\mathcal{L}_{\text{CE}}^{Y^{T_2}}$ are the cross-entropy losses for the binary change map and semantic maps at times $T_1$ and $T_2$, respectively. $\mathcal{L}_{\mathrm{SeK}}$ enforces semantic consistency within detected change regions. $\lambda_1$ and $\lambda_2$ are hyperparameters controlling the relative importance of the mIoU and SeK losses. Through experimentation, we selected $\lambda_1 = 0.15$ and $\lambda_2 = 0.3$.

\section{Experiments}\label{sec:exp}
\subsection{Datasets}\label{subsec:datasets}

In this study, we utilize two primary datasets for SCD. Below, we provide detailed descriptions of these datasets, emphasizing their distinct characteristics and application scenarios.

\textbf{SECOND Dataset~\cite{A_Yang_2022}}: The SECOND dataset is a semantic change detection dataset consisting of 4,662 pairs of aerial RGB images captured at a spatial resolution of 0.53 meters/pixel. Each image is of size 512 $\times$ 512 pixels. The dataset covers urban regions including Hangzhou, Chengdu, and Shanghai, with annotations across six key land cover classes: non-vegetated surfaces, trees, low vegetation, water bodies, buildings, and playgrounds. This diversity allows for comprehensive evaluation of semantic change detection methods across various urban dynamics. We adopt the standard split of 2,968 training pairs and 1,694 test pairs for our experiments.

\textbf{Landsat-SCD Dataset~\cite{A_Yuan_2022}
}: The Landsat-SCD dataset comprises 2,425 original RGB image pairs with a spatial resolution of 30 meters/pixel, captured over Tumushuke, Xinjiang, China, from 1990 to 2020. These images, each sized 416 $\times$ 416 pixels, are annotated for changes across four land cover classes, namely farmland, desert, buildings, and water bodies, associated with 10 specific semantic change types. The dataset is divided into training, validation, and testing sets with a ratio of 3:1:1, corresponding to 1,455, 485, and 485 samples, respectively.

\begin{table*}[htbp]
	\centering
	\caption{Quantitative results of the ablation study performed on the SECOND and Landsat-SCD datasets. Reported values are averaged over five independent runs.}
	\label{tab:ablation}
	\setlength{\tabcolsep}{4pt}
	\renewcommand{\arraystretch}{1.1}
	\begin{tabular}{@{}lccc cccc cccc@{}}
		\toprule
		& & & & \multicolumn{4}{c}{\textbf{SECOND}} & \multicolumn{4}{c}{\textbf{Landsat-SCD}} \\
		\cmidrule(lr){5-8}\cmidrule(lr){9-12}
		\textbf{Method} 
		& CGA & SeK & $F_i^{T_1},F_i^{T_2}$ 
		& OA (\%) & $F_{\text{scd}}$ (\%) & mIoU (\%) & SeK (\%)
		& OA (\%) & $F_{\text{scd}}$ (\%) & mIoU (\%) & SeK (\%) \\
		\midrule
		Base Model               & —          & —          & —          
		& 86.08 & 62.49 & 72.65 & 23.70 
		& 95.17 & 85.54 & 86.48 & 52.50 \\
		w/o CGA                  & —          & \checkmark & \checkmark 
		& 88.08 & 63.61 & 73.93 & 24.07 
		& 95.75 & 87.26 & 87.89 & 56.48 \\
		w/o SeK Loss             & \checkmark & —          & \checkmark 
		& 88.43 & 65.09 & 73.67 & 24.71 
		& 95.77 & 86.76 & 87.52 & 56.56 \\
		w/o FFT2 Branch          & \checkmark & \checkmark & —          
		& 87.86 & 64.52 & 73.16 & 24.03 
		& 96.06 & 88.72 & 88.27 & 58.75 \\
		w/o SeK Loss, w/o CGA    & —          & —          & \checkmark 
		& 87.26 & 63.30 & 72.89 & 23.42 
		& 95.68 & 87.58 & 87.30 & 55.96 \\
		\ours{}                  & \checkmark & \checkmark & \checkmark 
		& 88.62 & 65.78 & 74.07 & 25.50 
		& 96.25 & 89.27 & 88.81 & 60.26 \\
		\bottomrule
	\end{tabular}
\end{table*}

\subsection{Implementation Details}
Our architecture is implemented in PyTorch and trained using the AdamW~\cite{loshchilov2019decoupledweightdecayregularization}
optimizer with a learning rate of $1 \times 10^{-4}$, weight decay of $5 \times 10^{-3}$, and a batch size of 4. Training is conducted for 30{,}000 iterations on the SECOND dataset and 50{,}000 iterations on the Landsat-SCD dataset. To improve robustness against variations in illumination, viewpoint, and appearance, we apply extensive data augmentation, including random rotations, horizontal and vertical flips, and photometric adjustments to saturation, contrast, and brightness.

All experiments are performed on a workstation equipped with an NVIDIA RTX~A6000 GPU (48\,GB VRAM), a 32-core CPU, and 126\,GB of system memory. Under these settings, training requires approximately 24 hours for SECOND and 36 hours for Landsat-SCD. On this hardware, \ours{} records a peak GPU memory usage of 1152.65\,MB and an average inference time of 284.69\,ms per $512\times512$ bi-temporal RGB pair. For reproducibility, we fix the random seed to 42 for Python and PyTorch (including the data loader) in all experiments. The complete source code, configuration files, and pre-trained models are publicly released at \url{https://github.com/Buddhi19/Mamba-FCS.git} upon publication.

\subsection{Evaluation Metrics}\label{sec:metrics}

To assess the performance of our framework for SCD, we adopt a comprehensive set of metrics tailored to each task, widely utilized in the literature\cite{J_Ding_2024}
. We employ Overall Accuracy (OA), Mean Intersection over Union (mIoU), Separated Kappa (SeK), and the SCD-targeted F1 score (\( F_{\text{scd}} \)).

OA measures the proportion of correctly classified pixels relative to the total number of pixels, derived from the confusion matrix \( Q = \{ q_{ij} \} \), as described in Section~\ref{sec:loss}. This is formulated as,
	\begin{equation}
		\text{OA} = \frac{\sum_{i=1}^{N} q_{ii}}{\sum_{i=1}^{N} \sum_{j=1}^{N} q_{ij}}.
	\end{equation}
Due to the dominance of the no-change class in SCD tasks, OA alone may not fully capture performance in change regions. To address this mIoU and SeK are introduced \cite{A_Yang_2022} to evaluate the performance of CD and Semantic Exploitation (SE) respectively.

mIoU is computed as the average IoU across the no-change and change classes, with formulations for \( \mathrm{IoU}_1 \) and \( \mathrm{IoU}_2 \) provided in Equations \eqref{eq:iou1} and \eqref{eq:iou2}, respectively. SeK evaluates semantic agreement within changed regions, offering a targeted assessment of SCD performance. Its formulation is detailed in \ref{sec:SeK} , with the corresponding loss term defined in Equation \eqref{sek-eq}. \( F_{\text{scd}} \) quantifies segmentation accuracy in change regions, based on precision and recall for semantic classes with change annotations. This can be formulated as,
	\begin{equation}
		P_{\text{scd}} = \frac{\sum_{i=2}^{N} q_{ii}}{\sum_{i=2}^{N} \sum_{j=1}^{N} q_{ij}},
	\end{equation}
	\begin{equation}
		R_{\text{scd}} = \frac{\sum_{i=2}^{N} q_{ii}}{\sum_{i=1}^{N} \sum_{j=2}^{N} q_{ij}},
	\end{equation}
	\begin{equation}
		F_{\text{scd}} = \frac{2 \cdot P_{\text{scd}} \cdot R_{\text{scd}}}{P_{\text{scd}} + R_{\text{scd}}}.
	\end{equation}
\section{Results and Discussion}\label{sec:results}
This section presents the experimental results on benchmark datasets. We begin with quantitative ablation on the SECOND and Landsat-SCD datasets, and qualitative ablation on the SECOND dataset isolating the contributions of the Joint Spatio–Frequency Fusion mechanism, the Change-Guided Attention (CGA) module, and the SeK-inspired loss, as well as their combined effect. We then analyse the loss design itself by comparing the SeK-inspired term with alternative imbalance-aware losses and by studying the sensitivity of its weighting coefficients. Next, we benchmark Mamba-FCS against SOTA CNN, Transformer, and Mamba based SCD methods on both datasets, using both scalar metrics and qualitative visualisations, including dataset-level from–to confusion matrices and signed error maps to examine land-cover transition patterns. Finally, we report the computational cost and efficiency of Mamba-FCS relative to representative Transformer and Mamba baselines to assess its suitability for large-scale deployment.

\subsection{Ablation Studies on Key Contributions}
Our proposed architecture, \ours, comprises three key components: (i) a Joint Spatio–Frequency Fusion mechanism, (ii) a Change-Guided Attention (CGA) module, and (iii) a SeK-inspired loss. To isolate their individual contributions, we conduct ablation experiments on the SECOND and Landsat-SCD datasets by systematically removing each component. Specifically, we compare variants with and without CGA and with and without the SeK-inspired loss using both quantitative metrics and qualitative visualizations. Within the fusion mechanism, we further analyze the role of the FFT2 branch through combined quantitative and qualitative evaluation. Finally, we assess the joint impact of simultaneously disabling both the CGA module and the SeK-inspired loss, as well as removing all three components, to highlight their complementary benefits.
\begin{figure}
    \centering
    \includegraphics[width=1\linewidth]{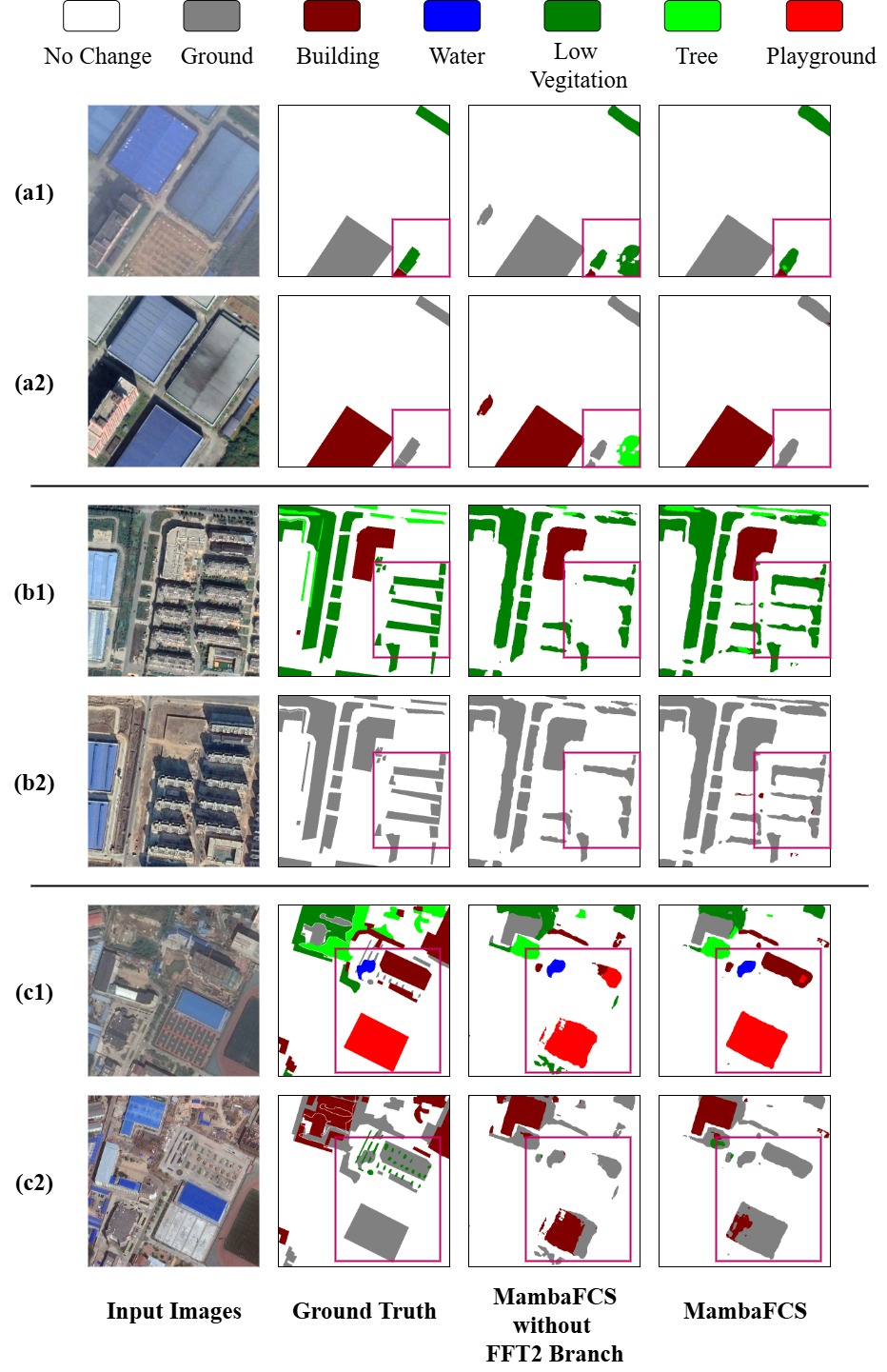}
    \caption{The qualitative impact of \emph{removing} the FFT2 branch. Columns (from left to right) display bi-temporal inputs, ground truth, results \emph{without} the FFT2 branch, and the full model. Red boxes highlight regions where frequency-domain cues enhance edge precision and suppress hallucinated changes.}
    \label{fig:fft_ablation}
\end{figure}

Quantitative results, are reported in Table~\ref{tab:ablation}. Qualitative analyses on SECOND dataset for the FFT2 branch, CGA, and Sek Inspired Loss are presented in Figures~\ref{fig:fft_ablation}, \ref{fig:CGA_ablation}, and \ref{fig:SeK_ablation}, respectively. Each figure is structured with four columns. The first displays input bi-temporal images, the second shows the Ground Truth (GT), the third presents the model’s output without the ablated component, and the fourth depicts the full model’s output. \textcolor{red}{Red} boxes highlight regions where each component significantly enhances performance. The results in Table~\ref{tab:ablation} and the qualitative visualizations collectively demonstrate that each component plays a valuable role in improving the model’s performance for SCD.

\subsection{Effect of FFT2 Branch}
Removing the FFT2 Branch results in performance drops for both benchmarks, as shown in Table~\ref{tab:ablation}. This underscores the FFT2 Branch's critical role in enhancing model performance. In Figure~\ref{fig:fft_ablation}, rows (a1) and (a2) demonstrate the FFT2 Branch’s ability to recover sub-pixel edges and suppress hallucinatory changes, which are prevalent without the FFT2 Branch. In rows (b1) and (b2), under challenging conditions such as heavy shadows and low contrast, the FFT2 Branch effectively detects transitions from low vegetation to ground. Rows (c1) and (c2) show FFT2 branch reduces boundary erosion and false changes in unchanged structures, alongside improved detection of changes from buildings to non-vegetated surfaces and reduced false detections of buildings on ground in row (f).

In summary, incorporating log-amplitude frequency features into the fusion block enables precise detection of fine-grained changes that spatial convolutions alone struggle to capture in this case. These qualitative improvements align with the quantitative gains in SeK, mIoU, $F_{scd}$, and OA.

\subsection{Effect of Change-Guided Attention (CGA)}
\begin{figure}
    \centering
    \includegraphics[width=1\linewidth]{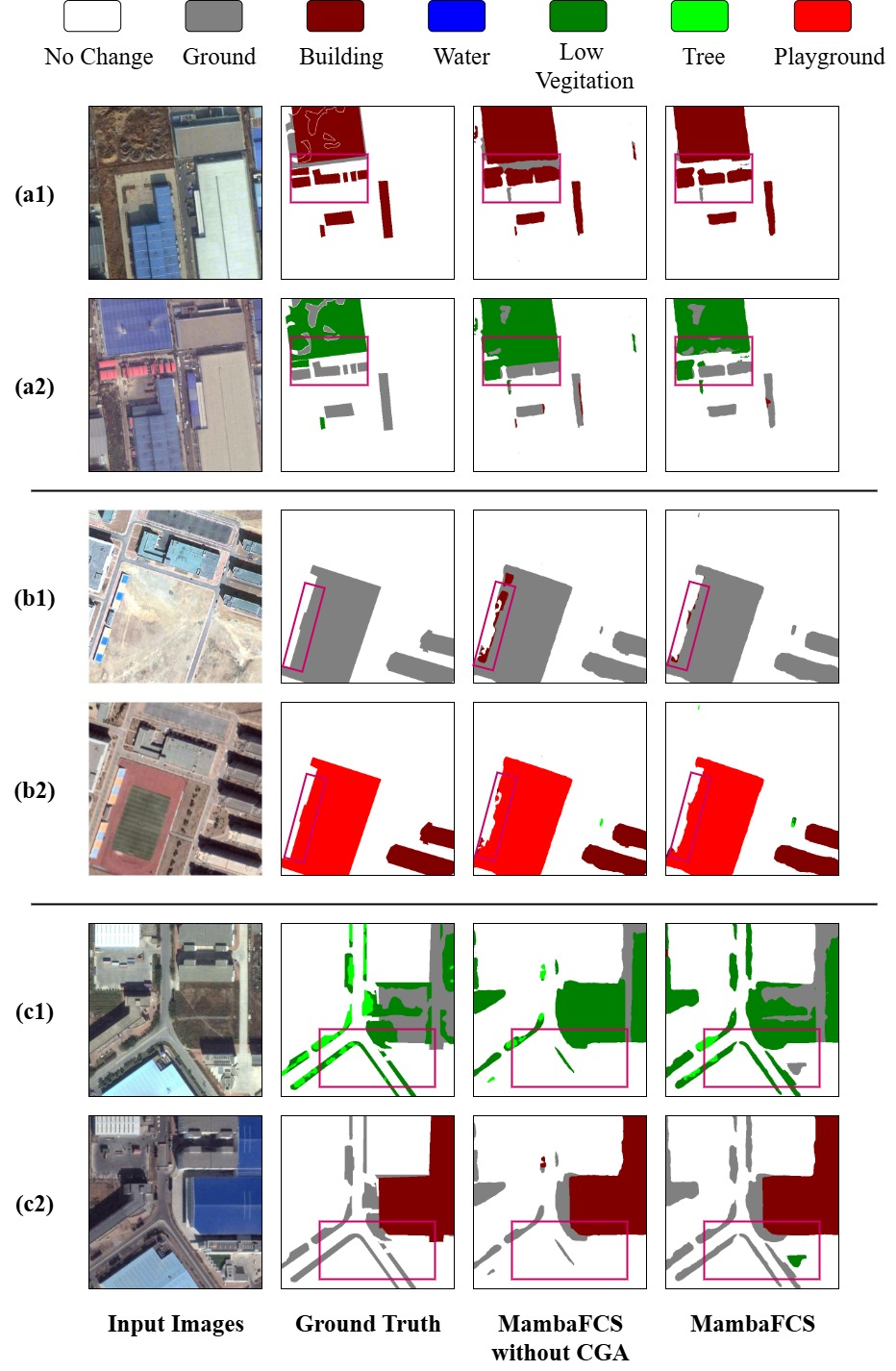}
    \caption{The qualitative impact of \emph{removing} the Change-Guided Attention (CGA) module. Columns (from left to right) display bi-temporal inputs, ground truth, results \emph{without} the CGA module, and the full model. Red boxes highlight regions where CGA sharpens change masks, reduces class confusion at boundaries, and mutually enhances both BCD and SCD performance.}
    \label{fig:CGA_ablation}
\end{figure}

Table~\ref{tab:ablation} shows that removing the CGA module leads to performance decreases of 1.43\%, 0.14\%, 2.17\%, and 0.54\% in SeK, mIoU, $F_{\text{scd}}$, and OA, respectively, for the SECOND dataset, and 3.78\%, 0.92\%, 2.01\%, and 0.50\% for the Landsat-SCD dataset. Figure~\ref{fig:CGA_ablation} further illustrates the pivotal role of CGA, which interconnects the three decoders in our architecture ($\mathcal{F}_{BCD}$, $\mathcal{F}^{T_1}_{SCD}$, and $\mathcal{F}^{T_2}_{SCD}$). Rows (a1) to (b2) demonstrate CGA’s ability to improve change map fidelity, and suppress false changes around changed areas. Rows (c1) and (c2) highlight CGA’s benefits for both BCD and SCD tasks, showing that without CGA, independent operation of decoders leads to fragmented change masks and mislabeled semantics at boundaries. By facilitating information exchange among decoders, CGA enhances coordinated learning, directly contributing to the quantitative gains observed in Table~\ref{tab:ablation}.

\begin{figure}
    \centering
    \includegraphics[width=1\linewidth]{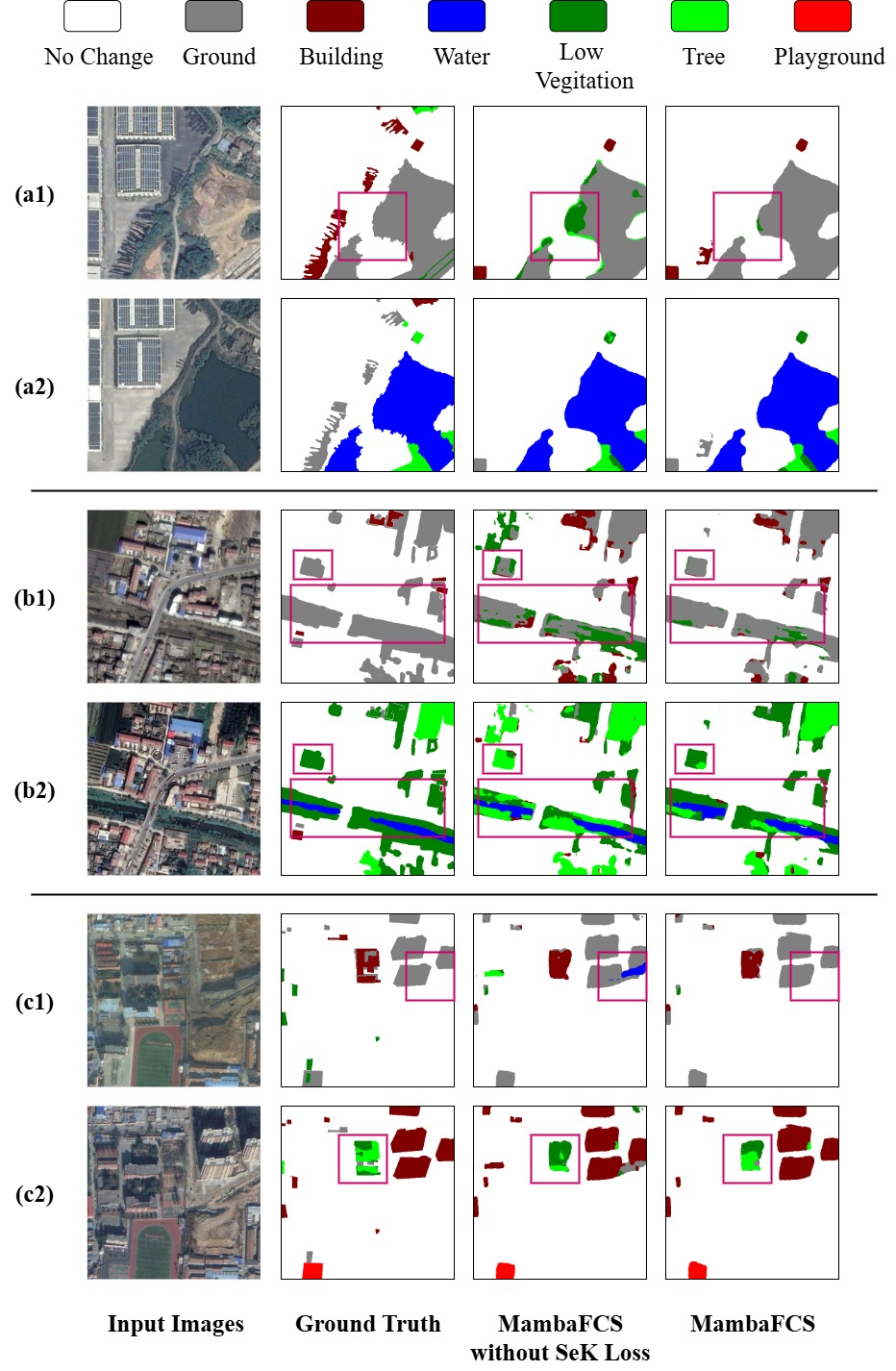}
    \caption{The qualitative impact of omitting the Separated Kappa (SeK) Loss. Columns (from left to right) display bi-temporal inputs, ground truth, results \emph{without} SeK, and the full model. Red boxes highlight regions where omitting SeK Loss increases false semantic labels, while including SeK reduces artifacts.}
    \label{fig:SeK_ablation}
\end{figure}

\subsection{Effect of SeK Inspired Loss}
Removing the SeK-inspired loss during training results in performance drops of 0.79\%, 0.40\%, 0.69\%, and 0.19\% in SeK, mIoU, $F_{\text{scd}}$, and OA, respectively, for the SECOND dataset, and 3.70\%, 1.29\%, 2.51\%, and 0.48\% for the Landsat-SCD dataset, as shown in Table~\ref{tab:ablation}. The most significant drop occurs in the SeK metric, which emphasizes class-balanced semantic change identification, confirming that Sek Inspired Loss primarily enhances semantic class accuracy. Figure~\ref{fig:SeK_ablation} provides qualitative evidence supporting this observation. Rows (a1) and (b1) illustrate how Sek Inspired Loss reduces false occurrences of low vegetation on ground. Similarly, rows (b2) and (c2) show its effectiveness in mitigating false occurrences of trees on low vegetation. Row (c1) demonstrates a complete elimination of false occurrences of water. Notably, Sek Inspired Loss enhances differentiation between changed and unchanged regions across all rows, slightly improving change map accuracy.

In summary, Sek Inspired Loss strengthens semantic class identification by prioritizing class-balanced change detection, reducing false changes and improving accuracy, as evidenced by quantitative metrics in Table~\ref{tab:ablation} and qualitative results in Figure~\ref{fig:SeK_ablation}.

\subsection{Combined Effect of CGA and SeK Inspired Loss}
Table~\ref{tab:ablation} shows that removing both the CGA module and Sek Inspired Loss results in the largest performance degradation, confirming their synergistic effect.

The Sek Inspired Loss directly supervises the semantic decoders, rewarding accurate semantics within change regions, but does not directly influence the Binary Change Decoder. The CGA module, however, connects the Binary Change Decoder to the Semantic Map Decoders, enabling indirect propagation of Sek Inspired Loss gradients to the Binary Change Decoder. Consequently, with CGA enabled, the Binary Change Decoder benefits from Sek Inspired Loss supervision, enhancing both change detection and semantic consistency. This interplay explains the significant performance drop when both components are removed, validating their combined use in \ours{}.

\subsection{Loss Function Analysis}
To justify our loss design for SCD under severe class imbalance, we conduct two complementary ablations on the \textsc{SECOND} dataset.

First, we compare three additional imbalance handling terms Focal\cite{DBLP:journals/corr/abs-1708-02002}, Dice\cite{Sudre_2017} and the proposed SeK inspired term while keeping the baseline supervision and the \ours{} architecture fixed. As summarised in Table~\ref{tab:loss_ablation}, the SeK inspired term achieves higher SeK and mIoU than both Focal and Dice losses, while OA and $F_{\text{scd}}$ remain very similar, indicating a more effective focus on under-represented semantic transitions.

\begin{table*}[t]
    \centering
    \caption{Effect of different additional loss terms for handling class imbalance on the \textsc{SECOND} dataset.}
    \label{tab:loss_ablation}
    \begin{tabular}{lccccccc}
    	\toprule
    	\multirow{2}{*}{Additional term} & \multicolumn{3}{c}{Loss components} & \multicolumn{4}{c}{\textsc{SECOND}} \\
    	\cmidrule(lr){2-4}\cmidrule(lr){5-8}
    	& Focal & Dice & SeK & OA (\%) & $F_{\text{scd}}$ (\%) & mIoU (\%) & SeK (\%) \\
    	\midrule
    	Focal               & $\checkmark$ & ---          & ---          & 88.23 & 65.36 & 73.90 & 24.90 \\
    	Dice                & ---          & $\checkmark$ & ---          & 88.42 & 64.93 & 73.86 & 24.81 \\
    	SeK-inspired (ours) & ---          & ---          & $\checkmark$ & \textbf{88.62} & \textbf{65.78} & \textbf{74.07} & \textbf{25.50} \\
    	\bottomrule
    \end{tabular}
\end{table*}

Second, we study the weighting coefficients $\lambda_1$ and $\lambda_2$ in the SeK-inspired loss. Table~\ref{tab:lambda_ablation} shows that extreme or unbalanced weights reduce SeK and mIoU, whereas the configuration $\lambda_1=0.15$ and $\lambda_2=0.30$ provided a reasonable balance across metrics and is used in our experiments to achieve the best performance.

{\color{red}
\begin{table}[t]
    \centering
    \caption{Sensitivity of the SeK-inspired loss to different choices of $\lambda_1$ and $\lambda_2$ on the \textsc{SECOND} dataset}
    \label{tab:lambda_ablation}
    \begin{tabular}{cc|cccc}
    	\toprule
    	$\lambda_1$ & $\lambda_2$ & OA (\%) & $F_{\text{scd}}$ (\%) & mIoU (\%) & SeK (\%) \\
    	\midrule
    	1.00  & 1.00  & 88.36 & 64.94 & 73.85 & 24.84 \\
    	0.50  & 1.00  & 88.58 & 65.15 & 73.98 & 25.03 \\
    	0.30  & 0.15  & 88.06 & 64.77 & 73.32 & 24.25 \\
    	\textbf{0.15}  & \textbf{0.30}  & \textbf{88.62} & \textbf{65.78} & \textbf{74.07} & \textbf{25.50} \\
    	\bottomrule
    \end{tabular}
\end{table}
}

\subsection{Comparative Experiments}

To assess our model’s performance, we benchmark against SOTA methods for SCD. The CNN‐based competitors include HRSCD variants (S2–S4)~\cite{M_Caye_2019}, ChangeMask~\cite{C_Zheng_2022}, SSCD-1 and Bi-SRNet~\cite{B_Ding_2022}, and TED~\cite{J_Ding_2024}. Transformer‐based baselines are SMNet~\cite{S_Niu_2023} and ScanNet~\cite{J_Ding_2024}. We also compare our model with Mamba-based baselines such as ChangeMamba~\cite{C_Chen_2024}.

\subsubsection{\textbf{Quantitative and Qualitative Results}}
\begin{figure*}
    \centering
    \includegraphics[width=0.85\linewidth]{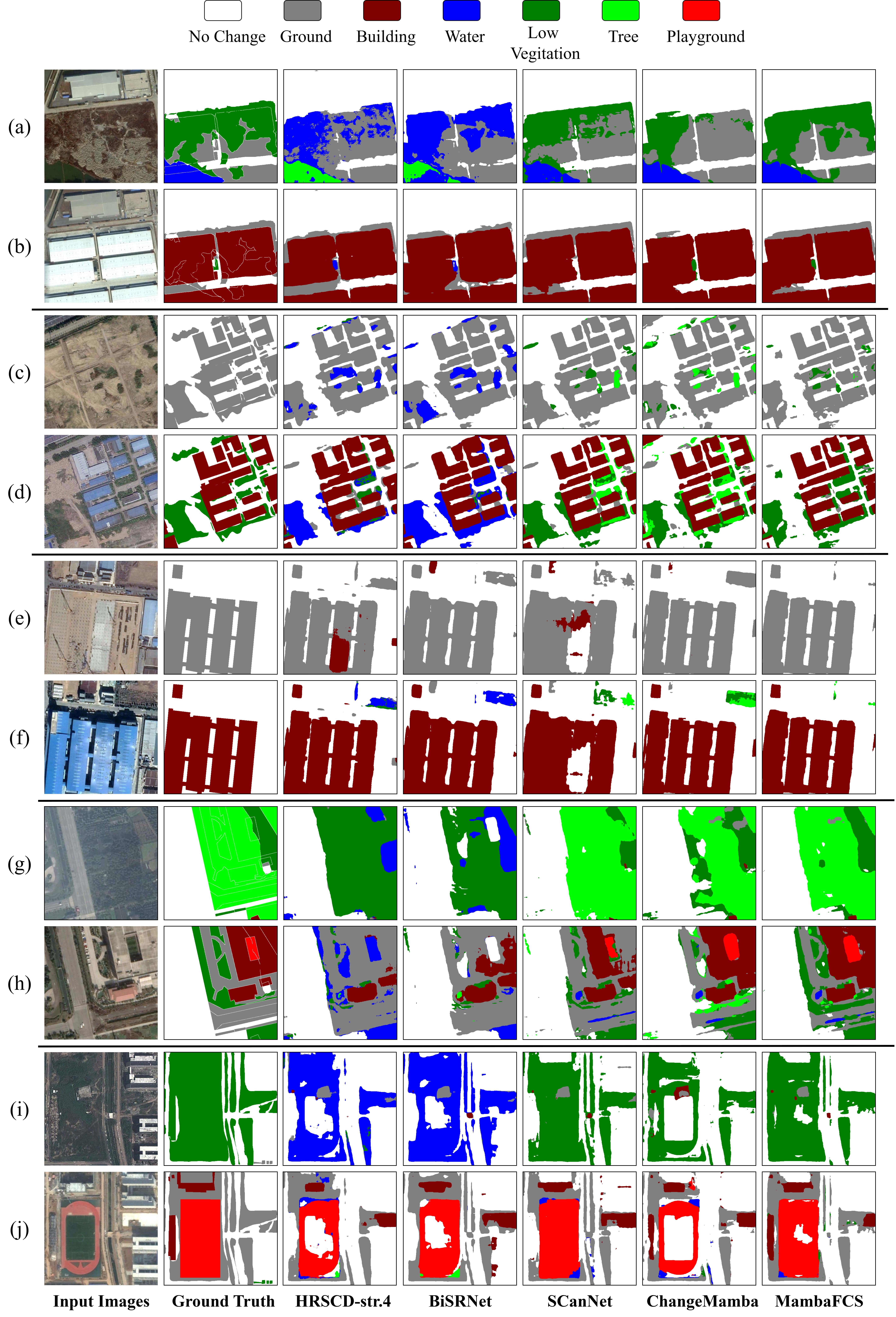}
    \caption{Qualitative comparison on the test split of \textbf{SECOND} dataset.  
    	Columns show (i) bi-temporal inputs, (ii) ground truth, and predictions from (iii) \emph{HRSCD-S4}, (iv) \emph{Bi-SRNet}, (v) \emph{SCanNet}, (vi) \emph{ChangeMamba}, and  (vii) \emph{Mamba-FCS}.}
    \label{fig:second_sota}
\end{figure*}

\begin{figure*}
    \centering
    \includegraphics[width=0.85\linewidth]{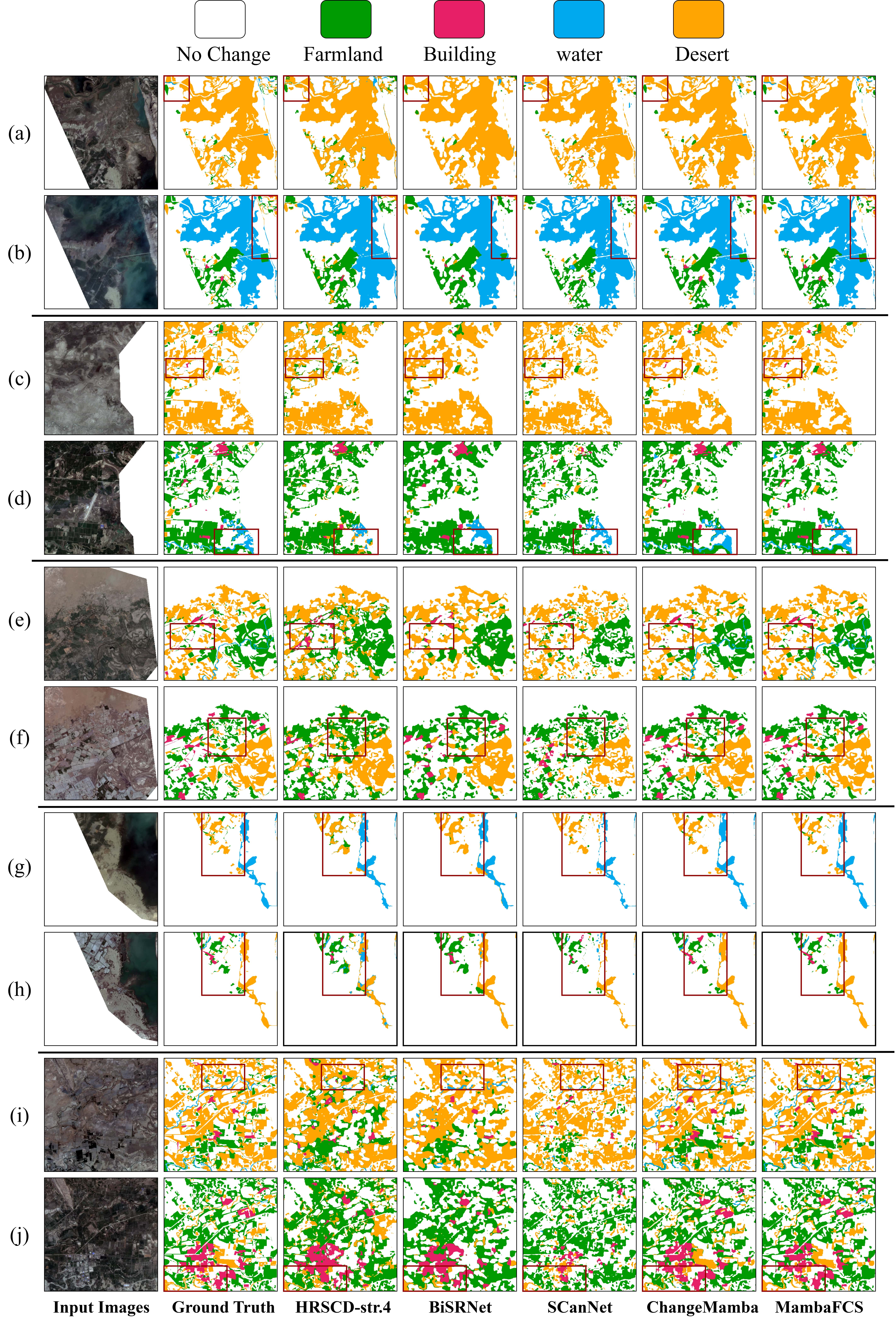}
    \caption{Qualitative comparison on the test split of  \textbf{Landsat-SCD} dataset. Columns display (i) bi-temporal inputs, (ii) ground truth, and predictions from (iii) \emph{HRSCD-S4}, (iv) \emph{Bi-SRNet}, (v) \emph{SCanNet}, (vi) \emph{ChangeMamba}, and  (vii) \emph{Mamba-FCS}. \textcolor{red}{Red} boxes
    	highlight regions where performance improvement is clearly visible.}
    \label{fig:landsat_sota}
\end{figure*}

We present quantitative results for the \textit{SECOND} dataset and for the \textit{Landsat-SCD} dataset in Table~\ref{tab:combined_sota}. For clarity, SOTA competitors are categorized into three families: (i) CNN-based models, (ii) Transformer-based models, and (iii) Mamba-based models. Qualitative comparisons are provided through segmentation maps in Fig.~\ref{fig:second_sota} for the SECOND dataset and Fig.~\ref{fig:landsat_sota} for the Landsat-SCD dataset.

\paragraph{SECOND Dataset}
Among CNN-based models, Bi-SRNet achieves the highest performance, with an OA of 87.84\%, an $F_{scd}$ score of 62.61\%, a mIoU of 73.41\%, and a SeK of 23.22\%. The lightweight TED model follows, demonstrating that its spatial-preserving design effectively captures semantic changes without extensive temporal reasoning. Within the Transformer family, ScanNet leverages long-range spatio-temporal dependencies, improving performance to 63.66\% $F_{scd}$ and 23.94\% SeK, the best results outside the Mamba family. By replacing quadratic self-attention with linear state-space mixing, ChangeMamba advances the frontier, achieving 88.12\% OA, 64.03\% $F_{scd}$, 73.68\% mIoU, and 24.11\% SeK.

Building on the same backbone and integrating the aforementioned techniques, \ours{} achieves the best performance among the evaluated methods, surpassing previous results across all four metrics, achieving 88.62\% OA, 65.78\% $F_{scd}$, 74.07\% mIoU, and 25.50\% SeK on the SECOND dataset.

These results are corroborated by the qualitative analysis in Fig.~\ref{fig:second_sota}. In rows (a1), (a2), (b2), (d1), and (e1), HRSCD-str4 and Bi-SRNet consistently fail to identify the correct semantic class.nWhile ScanNet and ChangeMamba generally recover the correct semantic classes, their predictions still exhibit scattered false change blobs around building footprints and vegetation, particularly along boundaries. In contrast, \ours{} produces more compact building and playground regions, suppresses false changes in surrounding low-vegetation and tree areas, and better preserves unchanged structures. This systematic reduction of boundary noise and rare false transitions is consistent with the observed improvements in $F_{scd}$ and SeK, and is aligned with the design of the FFT2 branch and CGA, which respectively sharpen high-frequency edges and enforce coherence between the binary change and semantic decoders.

\begin{table*}[t]
	\centering
	\caption{Semantic change-detection results on the \textbf{SECOND} and \textbf{Landsat-SCD} datasets (higher is better).  
		For each dataset, best, second-best and third-best scores are highlighted in
		\textcolor{red}{red}, \textcolor{blue}{blue} and \textcolor{green}{green}, respectively. Standard train-test splits were used for both datasets.}
	\label{tab:combined_sota}
	\setlength{\tabcolsep}{10.5pt}
	\renewcommand{\arraystretch}{1.1}
	\begin{tabular}{@{}l cccc cccc@{}}
		\toprule
		& \multicolumn{4}{c}{\textbf{SECOND}} & \multicolumn{4}{c}{\textbf{Landsat-SCD}} \\
		\cmidrule(lr){2-5}\cmidrule(lr){6-9}
		\textbf{Method} & OA & $F_{\text{scd}}$ & mIoU & SeK & OA & $F_{\text{scd}}$ & mIoU & SeK \\ \midrule
		\multicolumn{9}{c}{\textbf{CNN-based models}}\\
		HRSCD-S2\cite{M_Caye_2019}  & 85.49 & 49.22 & 64.43 & 10.69 & 86.06 & 36.52 & 74.92 &  2.89 \\
		HRSCD-S3\cite{M_Caye_2019}  & 84.62 & 51.62 & 66.33 & 11.97 & 91.47 & 75.86 & 79.79 & 35.57 \\
		HRSCD-S4\cite{M_Caye_2019}  & 86.62 & 58.21 & 71.15 & 18.80 & 92.17 & 77.37 & 81.07 & 38.09 \\
		ChangeMask\cite{C_Zheng_2022}& 86.93 & 59.74 & 71.46 & 19.50 & 92.93 & 79.74 & 81.46 & 40.50 \\
		SSCD-1\cite{B_Ding_2022}    & 87.19 & 61.22 & 72.60 & 21.86 & 93.20 & 80.53 & 81.89 & 41.77 \\
		Bi-SRNet\cite{B_Ding_2022}  & 87.84 & 62.61 & 73.41 & 23.22 & 93.80 & 82.01 & 82.94 & 44.27 \\
		TED\cite{J_Ding_2024}       & 87.39 & 60.34 & 72.79 & 22.17 & 94.39 & 83.63 & 84.79 & 48.33 \\ \midrule
		\multicolumn{9}{c}{\textbf{Transformer-based models}}\\
		SMNet\cite{S_Niu_2023}      & 86.68 & 60.34 & 71.95 & 20.29 & {94.53} & {84.12} & {85.65} & {51.14} \\
		SCanNet\cite{J_Ding_2024}   & \textcolor{green}{87.86} & \textcolor{green}{63.66} & \textcolor{green}{73.42} & \textcolor{green}{23.94} & \textcolor{green}{96.04} & \textcolor{green}{85.62} & \textcolor{green}{86.37} & \textcolor{green}{52.63} \\ \midrule
		\multicolumn{9}{c}{\textbf{Mamba-based models}}\\
		ChangeMamba\cite{C_Chen_2024}
         & \textcolor{blue}{88.12} & \textcolor{blue}{64.03} & \textcolor{blue}{73.68} & \textcolor{blue}{24.11} & \textcolor{blue}{96.08} & \textcolor{blue}{86.61} & \textcolor{blue}{86.91} & \textcolor{blue}{53.66} \\
		\ours{}                     & \textcolor{red}{88.62} & \textcolor{red}{65.78} & \textcolor{red}{74.07} & \textcolor{red}{25.50} & \textcolor{red}{96.25} & \textcolor{red}{89.27} & \textcolor{red}{88.81} & \textcolor{red}{60.26} \\
		\bottomrule
	\end{tabular}
\end{table*}

\paragraph{Landsat-SCD Dataset}
Among CNN-based baselines, TED leads with 94.39\% OA, 83.63\% $F_{\text{scd}}$, 84.79\% mIoU, and 48.33\% SeK. Its lightweight encoder-decoder architecture demonstrates that preserving spatial details effectively captures temporal semantics in medium-resolution Landsat imagery. In the Transformer family, ScanNet, achieves 96.04\% OA, 85.62\% $F_{\text{scd}}$, 86.37\% mIoU, and 52.63\% SeK, the best results outside the Mamba family. ChangeMamba further improves performance to 96.08\% OA, 86.61\% $F_{\text{scd}}$, 86.91\% mIoU, and 53.66\% SeK.\ours{} achieves the best results among the methods evaluated in our experiments, achieving 96.25\% OA, 89.27\% $F_{\text{scd}}$, 88.81\% mIoU, and 60.26\% SeK on the Landsat-SCD dataset.

These findings are supported by the qualitative analysis in Fig.~\ref{fig:landsat_sota}. In rows (a1) and (a2), \ours{} captures water regions more accurately than competing models. Similarly, in rows (b1) to (e2), \ours{} substantially reduces these false alarms, yielding cleaner desert extents and more coherent farmland and water patches, particularly along transition fronts. This behaviour matches the quantitative gains in SeK, where \ours{} better preserves the distribution of major transitions while reducing noise for rare ones. For readers’ convenience, we highlight regions with noticeable improvements in \textcolor{red}{red}.

\begin{figure*}
	\centering
	\includegraphics[width=0.99\linewidth]{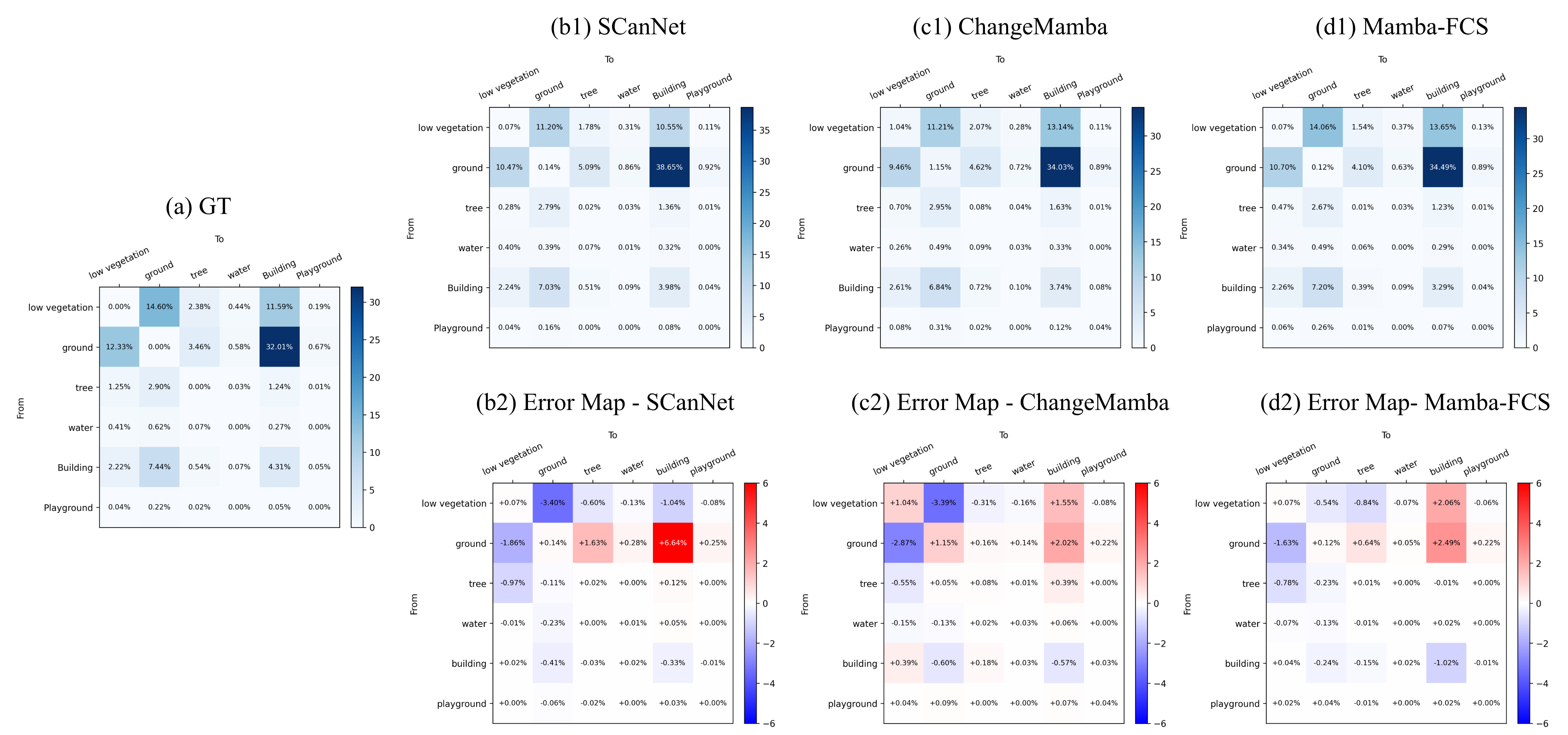}
	\caption{Dataset-level \emph{from–to} confusion matrices and corresponding error maps for the SECOND test split. The top row shows the percentage of changed pixels for each land-cover transition for gronud truth (a), SCanNet (b1), ChangeMamba (c1), and Mamba-FCS (d1); the bottom row (b2–d2) shows signed differences.}
	\label{fig:changeanalysis}
\end{figure*}

\subsubsection{\textbf{Change Analysis}}
\begin{figure*}
	\centering
	\includegraphics[width=0.99\linewidth]{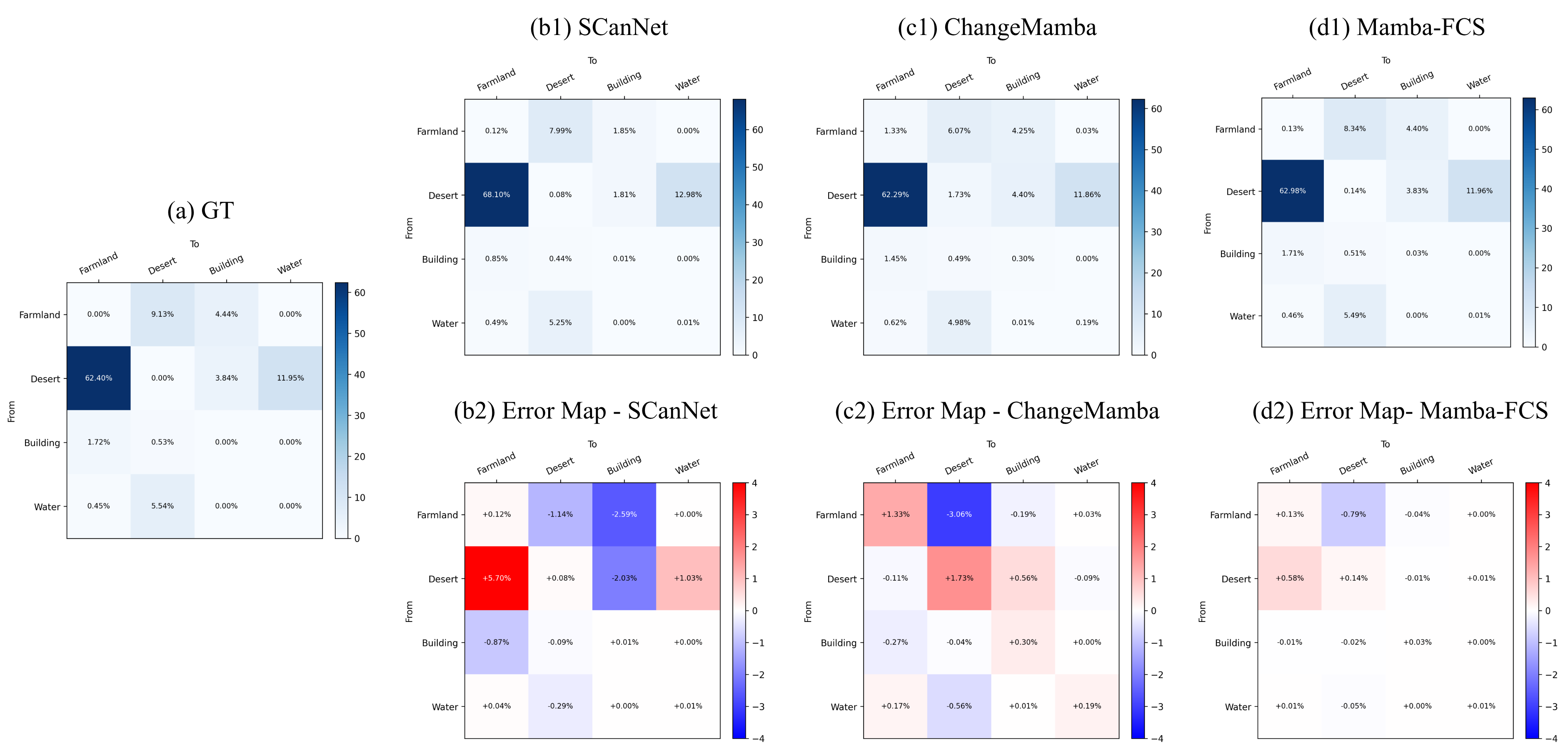}
	\caption{Dataset-level \emph{from–to} confusion matrices and corresponding error maps for the Landsat-SCD test split. The top row shows the percentage of changed pixels for each land-cover transition for ground truth (a), SCanNet (b1), ChangeMamba (c1), and Mamba-FCS (d1); the bottom row (b2–d2) shows signed differences}
	\label{fig:changeanalysis_landsat}
\end{figure*}

To provide a more interpretable view of how each model behaves on different land-cover transitions, we plot dataset-level ``from $\rightarrow$ to'' confusion matrices and corresponding signed error maps for the SECOND and LandSat-SCD test splits in Figures~\ref{fig:changeanalysis} and~\ref{fig:changeanalysis_landsat}. Each confusion matrix in the top row summarises the percentage of changed pixels moving from a source class (rows) to a target class (columns) for the ground truth and for the predictions of SCanNet, ChangeMamba, and \ours{}. The error maps in the bottom row then visualise, for every transition, the difference between the model prediction and ground truth, where red cells indicate overestimation and blue cells indicate underestimation.

On the \textsc{SECOND} dataset (Figure~\ref{fig:changeanalysis}), the ground-truth matrix is dominated by transitions such as \textit{ground} $\rightarrow$ \textit{building} and \textit{low vegetation} $\rightarrow$ \textit{ground}/\textit{building}, while transitions involving \textit{tree}, \textit{water}, and \textit{playground} are comparatively rare. SCanNet’s error map (b2) shows a strong positive bias on the \textit{ground} $\rightarrow$ \textit{building} entry and noticeable negative deviations on several other entries, revealing a tendency to overestimate urban expansion at the expense of other change patterns. ChangeMamba (c2) reduces some of these extremes but still exhibits distinct red/blue patches, particularly on transitions towards the \textit{building} class. In contrast, the error map of \ours{} (d2) is largely composed of light colours with only small residual deviations, indicating that the predicted distribution of land-cover conversions closely follows the ground truth. Rare transitions in the lower-right region of the matrix remain close to zero in both value and error, showing that \ours{} avoids introducing spurious changes for infrequent categories.

A similar trend appears on the LandSat-SCD dataset (Figure \ref{fig:changeanalysis_landsat}). Here, the ground truth is dominated by transitions such as \textit{desert} $\rightarrow$ \textit{farmland} and \textit{desert} $\rightarrow$ \textit{water}, with other conversions occurring much less frequently. SCanNet’s error map (b2) contains pronounced red regions on \textit{desert} $\rightarrow$ \textit{farmland} and related entries, indicating systematic overestimation of agricultural expansion, while ChangeMamba (c2) still shows visible positive and negative patches across several desert-related transitions. In contrast, the error map of \ours{} (d2) is almost uniformly close to zero, with only mild residual deviations on the dominant transitions and very small errors on rare ones, such as conversions between \textit{building} and \textit{water}.

Overall, these confusion matrices and signed error maps confirm that \ours{} not only improves scalar metrics such as SeK and mIoU, but also reduces systematic biases in specific land-cover conversions and better preserves the empirical distribution of both common and rare transitions. This behaviour is especially desirable for environmental monitoring and urban-growth analysis, where over- or under-estimating particular transitions (e.g., urbanisation or farmland expansion) could lead to misleading conclusions about the underlying dynamics.

\subsubsection{\textbf{Computational Cost and Efficiency}}

The table \ref{tab:scd_cost} compares the computational cost for Mamba Based methods with transformer based models. Although \ours{} has the largest number of parameters among the compared methods, its computational cost remains on par with the Transformer baseline ScanNet (263.15 vs. 264.95 GFLOPs) despite having roughly 7 times more parameters, and only moderately higher than ChangeMamba (211.55 GFLOPs) while using more than 2 times the parameters. This decoupling between parameter count and FLOP cost stems from the linear-time state-space operations in VMamba, which avoid the quadratic scaling of self-attention with spatial resolution. 

In practice, this means that \ours{} can afford a higher-capacity backbone and richer decoders without incurring a proportional increase in inference-time computation, making it more suitable for large-scale, high-resolution SCD deployments.

\begin{table}[t]
  \centering
  \caption{Comparison of best performing SCD models in computational cost (input: 512\,$\times$\,512 two-epoch RGB).}
  \label{tab:scd_cost}
  \setlength{\tabcolsep}{6pt}
  \renewcommand{\arraystretch}{1.1}
  \begin{tabular}{@{}l c c@{}}
    \toprule
    \textbf{Method} & \textbf{Params (M)} & \textbf{GFLOPs} \\
    \midrule
    \multicolumn{3}{c}{\textbf{Transformer-based models}} \\
    ScanNet~\cite{J_Ding_2024}   & 27.90 & 264.95 \\
    \midrule
    \multicolumn{3}{c}{\textbf{Mamba-based models}} \\
    ChangeMamba~\cite{C_Chen_2024}                & 89.99 & 211.55 \\
    \ours{}                      &   189.54     &  263.15       \\
    \bottomrule
  \end{tabular}
\end{table}

\section{Conclusions}\label{sec:conclusion}
In this paper, we presented \ours{}, a SCD framework specifically designed to effectively address the challenges of capturing long-range contextual dependencies and detecting subtle semantic transitions in high-resolution remote sensing imagery. 

Our approach combines the efficiency and global contextual modeling capabilities of VMamba, a linear-complexity visual state-space backbone, with a novel Joint Spatio-Frequency Feature Fusion strategy that incorporates log-amplitude frequency domain features to mitigate illumination-related artifacts and enhance fine-grained boundary detection. Additionally, we integrated a lightweight Change-Guided Attention (CGA) mechanism, which aligns semantic prediction heads with binary change cues, thereby improving semantic accuracy. Furthermore, we proposed the Separated Kappa(SeK) inspired loss function, re-purposing the established SCD evaluation metric into an effective training objective, particularly benefiting minority class transitions.

Extensive experiments on the SECOND and Landsat-SCD benchmarks demonstrate that our method outperforms recent SOTA approaches across multiple metrics, including overall accuracy, F1-score, mean Intersection over Union (mIoU), and Separability and Kappa (SeK). Specifically, our method achieves a SeK score of 25.50\% on the SECOND dataset and 60.26\% on the Landsat-SCD dataset. These results highlight its capability in detecting rare and subtle semantic transitions, with reduced false positives and enhanced class-wise accuracy. These results underscore the efficacy of \ours{} in addressing complex SCD challenges in remote sensing.

However, the SeK-inspired loss is defined under fully supervised, pixel-wise annotations, and its behavior under label noise or in weakly/semi-supervised regimes has yet to be explored. Future work will therefore focus on developing lighter or distilled variants of \ours{} for efficient deployment, extending the joint spatio–frequency fusion to multi-modal inputs, and adapting the SeK-guided training strategy to semi-supervised and domain-adaptive SCD scenarios.

\bibliographystyle{unsrt}
\bibliography{references}

\end{document}